\documentclass[article]{aa} 

\usepackage{graphicx}
\usepackage{txfonts}
\usepackage{natbib}
\bibpunct{(}{)}{;}{a}{}{,}
\usepackage{amssymb}
\usepackage{xfrac}
\usepackage{xcolor}

\usepackage{hyperref}

\definecolor{red}{HTML}{E24A33}
\definecolor{blue}{HTML}{348ABD}
\definecolor{purple}{HTML}{988ED5}
\definecolor{gray}{HTML}{777777}
\definecolor{yellow}{HTML}{FBC15E}
\definecolor{green}{HTML}{8EBA42}
\definecolor{green2}{HTML}{5A8F00}
\definecolor{pink}{HTML}{FFB5B8}

\newcommand{\figref}[1]{Fig.~\ref{#1}} 

\newcommand{\tabref}[1]{Table~\ref{#1}}

\newcommand{\degree}{$^{\circ}$}
\renewcommand{\deg}{^{\circ}}

\newcommand{\Msun}{$M_{\rm\sun}$}

\newcommand{\klam}{$k\lambda$}
 
\newcommand{\dlq}[1]{``}
\newcommand{\drq}[1]{''}

\newcommand{\err}[2]{$\substack{+#1\\-#2}$}

\newcommand{\uv}{$\mathit{u}$-$\mathit{v}$}

\begin{document}

   \title{Constraining the physical structure of the inner few 100 AU scales of deeply embedded low-mass protostars.\thanks{Based on observations carried out with the IRAM Plateau de Bure Interferometer. IRAM is supported by INSU/CNRS (France), MPG (Germany) and IGN (Spain).}}

   \author{M.~V. Persson
             \inst{1}\fnmsep
             \and D. Harsono\inst{1}\fnmsep\inst{2}
             \and J.~J. Tobin\inst{1}
             \and E.~F. van Dishoeck\inst{1}\fnmsep\inst{3}
                        \and J.~K. J{\o}rgensen\inst{4}\fnmsep\inst{5}
                        \and N. Murillo\inst{3}
                        \and S.-P. Lai\inst{6}\fnmsep\inst{7}
          }

   \institute{Leiden Observatory, Leiden University, P.O. Box 9513, 2300 RA Leiden, The Netherlands              
                        \email{magnusp@strw.leidenuniv.nl}
            \and
            Universit\"{a}t Heidelberg, Zentrum f\"{u}r Astronomie, Institut für Theoretische Astrophysik (ITA), Albert-Ueberle-Str. 2, 69120, Heidelberg, Germany
            \and 
            Max-Planck Institute f\"{u}r extraterrestrische Physik (MPE), Giessenbachstrasse, 85748 Garching, Germany
                \and 
                Centre for Star and Planet Formation, Natural History Museum of Denmark,
                University of Copenhagen, {\O}ster Voldgade 5-7, \\1350, K{\o}benhavn~K, Denmark
                \and
                Niels Bohr Institute, University of Copenhagen, Juliane
                Maries Vej 30, 2100 K{\o}benhavn~{\O}, Denmark
            \and
            Institute of Astronomy and Department of Physics, National Tsing Hua
            University, 101 Section 2 Kuang Fu Road, 30013 Hsinchu, Taiwan
            \and
            Academia Sinica Institute of Astronomy and Astrophysics, PO Box
            23-141, 10617 Taipei, Taiwan
                }

   \date{Received XXXX YY, 2015; accepted XXXX YY, 2016}

 
  \abstract
        {The physical structure of deeply embedded low-mass protostars (Class~0) on 
        scales of less than 300 AU is still poorly constrained. While molecular line 
        observations demonstrate the presence of disks with Keplerian rotation 
        toward a handful of sources, others show no hint of rotation. Determining 
        the structure on small scales (a few 100~AU) is crucial for 
        understanding the physical and chemical evolution from cores to disks.}
        {We determine the presence and characteristics of compact, 
        disk-like structures in deeply embedded low-mass protostars. A related goal 
        is  investigating how the derived structure affects the determination of 
        gas-phase molecular abundances on hot-core scales.}
   {Two models of the emission, a Gaussian disk intensity distribution and a 
   parametrized power-law disk model, are fitted to subarcsecond resolution 
   interferometric continuum observations of five Class~0 sources, including one 
   source with a confirmed Keplerian disk. Prior to fitting the models to the 
   de-projected real visibilities, the estimated envelope from an 
   independent model and any companion sources are subtracted. For reference, a 
   spherically symmetric single power-law envelope is fitted to the 
   larger scale emission ($\sim$1000~AU)  and investigated further for one of 
   the sources on smaller scales.}
   {The radii of the fitted disk-like structures range from $\sim 90-170$~AU, 
   and the derived masses depend on the method. Using the Gaussian disk model results in masses of $54-556\times10^{-3}$~\Msun, and using the power-law disk model gives $9-140\times10^{-3}$~\Msun. While the disk radii agree with previous estimates the masses are different for some of the sources studied. 
   Assuming a typical temperature distribution ($r^{-0.5}$), the fractional 
   amount of mass in the disk above 100~K varies from 7\% to 
   30\%.}
   {A thin disk model can approximate the emission and physical 
   structure in the inner few 100 AU scales of the studied deeply  embedded 
   low-mass protostars and paves the way for analysis of a larger sample with 
   ALMA. Kinematic data are needed to determine the presence of any 
   Keplerian disk. Using previous observations of p-H$_2^{18}$O, we 
   estimate the relative gas phase water abundances relative to total 
   warm H$_2$ to be $6.2\times10^{-5}$ (IRAS~2A), $0.33\times10^{-5}$ 
   (IRAS~4A-NW), $1.8\times10^{-7}$ (IRAS~4B), and $<2\times10^{-7}$ 
   (IRAS~4A-SE), roughly an order of magnitude higher than previously inferred 
   when both warm and cold H$_2$ were used as reference. A spherically symmetric single power-law envelope model fails to simultaneously   reproduce both the small- and large-scale emission.}
   

   \maketitle
%

\section{Introduction}

A gravitationally collapsing core of gas and dust marks the beginning of the 
star formation process in molecular clouds. These infall-dominated 
\textit{pre-stellar cores} evolve into envelope dominated, \textit{Class~0} 
sources with well-collimated bipolar outflows \citep{andre00}, hinting at the 
presence of a disk. Once the accretion has progressed enough, the source is 
classified as a disk dominated \textit{Class~I} source, characterized by a more 
tenuous envelope and a rotating circumstellar disk 
\citep[e.g.,][]{jorgensen09,beckwith89,beckwith93,brogan15}. In the subsequent 
\textit{Class~II} stage (T~Tauri source), the envelope has almost completely 
dissipated and the disk is characterized by Keplerian rotation and may 
have developed cavities 
\citep{guilloteau94,qi03,simon00,williams11,espaillat14}. Although there is 
evidence for disks at all stages of protostellar evolution, they have not yet  been 
characterized in the earliest stages. Early simulations of collapsing rotating 
clouds show that large-scale ($\sim1000$~AU) flattened infalling envelopes with 
small-scale ($\sim 100$~AU) disks are expected to form 
\citep[e.g.,][]{cassen81,terebey84,galli93b}. Tracing the evolution of 
disks through the entire star formation process will make it possible to 
determine when and how disks form and evolve. The physical characteristics of 
this early structure therefore mark the starting point of disk evolution 
studies. 

Interferometric observations of deeply embedded low-mass protostars taken with a resolution of  few arcseconds  have revealed unresolved compact components on scales $<300$~AU \citep[e.g.,][]{hogerheijde99,looney03,harvey03,jorgensen05b}, and these structures  are not  fit well by standard envelope models.
Rotating disks with sizes of $R<50\sim200$~AU have been found for a few Class~0 
sources based on kinematical analysis on the molecular lines 
\citep[e.g.,][]{tobin12,murillo13b,ohashi14,lindberg14,lee14}. This raises the 
questions of how early such disks form after collapse and how common they are.  High-resolution C$^{18}$O observations toward other Class~0 sources 
show velocity gradients perpendicular to the outflow toward some sources 
\citep{yen15}, but not necessarily in a rotationally supported disk 
\citep{brinch09,maret14}. Thus, the physical conditions  (i.e., density, 
temperature, mass) of deeply embedded low-mass protostars on small scales are poorly 
constrained.

The density and temperature structure on small scales also affect the
chemistry and derived abundances. Some low-mass protostars show strong lines of 
complex organic molecules that are thought to originate from regions where the 
dust temperature exceeds 100~K and all ices have been sublimated
\citep[e.g.,][]{vandishoeck98,bottinelli04,herbst09,oberg14}. These
``hot core'' or ``hot corino'' regions are small  (radii less than 100
AU), and abundances with respect to H$_2$ are often inferred through
comparison with a spherically symmetric power-law envelope structure,
with all material inside the 100~K radius assumed to contain
sublimated ice material. However, in a disk-like structure some
fraction of this dust on a small scale may be much cooler than 100~K
\citep[e.g.,][]{harsono14,harsono15b}. This means that the abundances of
species that are only thought to exist in regions $>$100 K may have
been underestimated.  A prominent example is water itself: lines of
H$_2^{18}$O have been imaged in low-mass protostars on subarcsecond
scales but inferred hot core abundances are more than an order of
magnitude lower than expected from water ice sublimation
\citep[e.g.,][]{jorgensen10a,persson12,persson13,visser13}. Water, in
turn, controls available reaction routes for other species \citep[e.g.,
H$_2$O destroying HCO$^+$;][]{jorgensen13}. 

The temperature and density structures of the envelope on larger
scales (several 1000~AU) is relatively well characterized through
modeling of the spectral energy distribution and the single-dish
submillimeter continuum spatial structure 
\citep[e.g.,][]{jorgensen02,kristensen12} and through extinction
mapping at infrared wavelengths \citep[e.g.,][]{tobin10}.  To what
extent continuum observations can be used to probe the small-scale
structures, and the related transition from the large-scale envelope,
is not yet clear.

With the increased sensitivity of interferometric data, it
is now possible to routinely obtain subarcsecond resolution continuum
interferometric observations at mm wavelengths. Here we  analyze such
data for five deeply embedded low-mass protostars (Class~0), including
one source for which the presence of a Keplerian rotating disk has
been established from kinematic data. After removing any large-scale
envelope contribution, two different models are fitted to the
de-projected binned real visibility amplitudes, a Gaussian intensity 
distribution, and a parametrized disk model. Although this analysis cannot 
determine whether rotationally supported disks are present, it can constrain the
basic parameters of any flattened, compact disk-like structure such as
radius and mass. The method presented here is intended as a proof of
concept that such structures can indeed be characterized and that
they have consequences for inferred abundances of species such as
water. These methods can subsequently be extended and applied to
current/future ALMA observations of a large sample of low-mass protostars and
to additional molecules.

%
\section{Sample and observations}\label{sec:sample}
Five sources were considered in this study, the four protostars
IRAS~2A, IRAS~4A-NW, IRAS~4A-SE, and IRAS~4B in the NGC~1333 star
forming region \citep[235~pc,][]{hirota08} and VLA~1623 in the
$\rho$~Ophiuchus star forming region \citep[120~pc,
][]{loinard08}. The last source has a confirmed Keplerian disk with
a radius of $\sim$150 AU as inferred from high-resolution C$^{18}$O data
\citep{murillo13b}.

Observational data come from several programs.\footnote{Continuum data 
for the sources are available through \url{http://vilhelm.nu} and at CDS via anonymous ftp to \url{cdsarc.u-strasbg.fr} (130.79.128.5)
or via \url{http://cdsweb.u-strasbg.fr/cgi-bin/qcat?J/A+A/}.} The 
continuum toward NGC~1333 IRAS~2A and 4B at 203.4~GHz has been obtained with the
Plateau de Bure Interferometer (PdBI\footnote{Now the NOrthern 
Extended Millimeter Array (NOEMA)}) by \citet{jorgensen10a}
and \citet{persson12}. One track ($\sim$8~hours) of complimentary A
configuration (extended) data were obtained for IRAS~4B on 2~February
2011. Phase and amplitude were calibrated on the quasar 0333+321, and
the absolute flux on the blazar 3C454.3. The calibration and imaging
was done in IRAM GILDAS. The observations of IRAS~4B cover (projected)
baselines from 12--515~\klam\ (21\arcsec--0\farcs5), and for IRAS~2A they cover 
10.5--307~\klam\ (23\arcsec-0\farcs8).
IRAS~4A was observed with the Submillimeter Array (SMA) at 335~GHz
(Lai et al. in prep). The observations of IRAS~4A cover
baselines from 10--510~\klam\ (25\arcsec--0\farcs5). Finally, the ALMA Cycle 2
VLA~1623 data are from \citet{murillo15} at 219.0~GHz and cover
baselines from 13.8--796.6~\klam\ (18\arcsec--0\farcs3). The various
source parameters are summarized in Table~\ref{tab:sources} and the 
sensitivity, beam, and peak fluxes for the observations are listed in 
\tabref{tab:obs}.

\begin{table}[ht]
\caption{Parameters for the studied sources.}\label{tab:sources}
\centering
\begin{tabular}{l c c c c c c}
\hline\hline          
        Source          & $L_\mathrm{bol}$ & $M_\mathrm{env}$  & $T_\mathrm{bol}$ & 
        $d$
        \\      
                                & ($L_\odot$) & ($M_\odot$) & ($K$)     & (pc) \\      
\hline                  
   IRAS 2A              & $35.7$        & $5.1$         & $50$ & 235 \\
   IRAS 4A\tablefootmark{a}             & $9.1$ & $5.6$ & $33$ & 235  \\
   IRAS 4B              & $4.4$         & $3.0$         & $28$   & 235 \\
   VLA 1623~A   & $1.1$         & $0.22$        & $10$   & 120 \\
\hline
\end{tabular}
\tablefoot{From \citet[][and references 
therein]{kristensen12,murillo13a,murillo13b}.  Distances are taken from 
\citet{hirota08} and \citet{loinard08}. 
\tablefoottext{a}{Values refer to both components of the binary (SE and NW).}
}
\end{table}

\begin{figure}[htp]
        \centering
        \includegraphics[width=\linewidth]{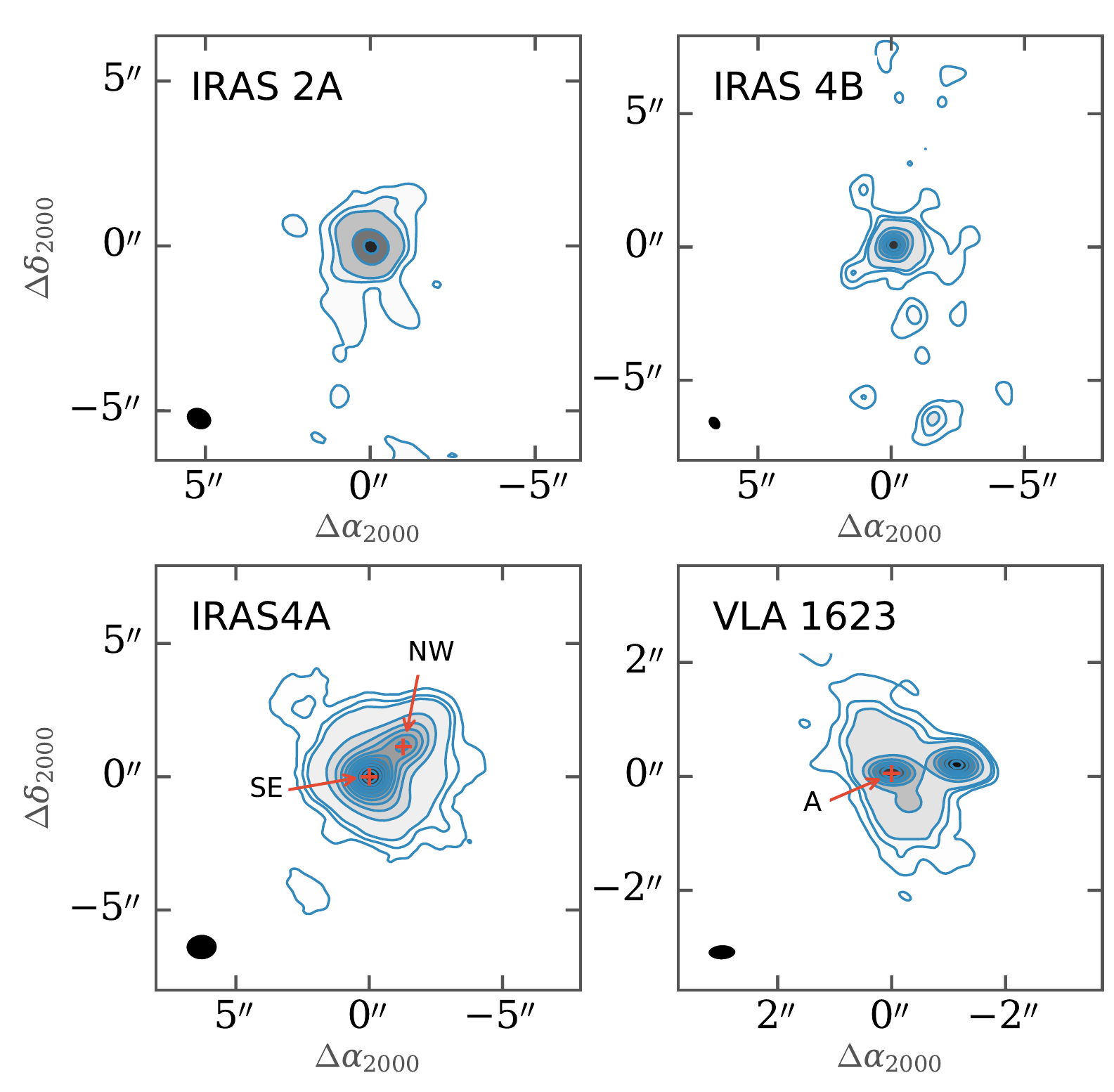}
        \caption{Maps of the continuum toward IRAS~2A (PdBI), 
        IRAS~4B (PdBI), IRAS~4A (SMA), and VLA~1623 (ALMA). Contours 
        start at 3$\sigma_\mathrm{RMS}$ in steps of 3$\sigma_\mathrm{RMS}$ until 
        9$\sigma_\mathrm{RMS}$, then in steps of 20$\sigma_\mathrm{RMS}$ 
        ($\sigma_\mathrm{RMS}$ in \tabref{tab:obs}). We note the 
        absence of      any 2\arcsec\ triple toward IRAS~2A. The observations of 
        IRAS 4B are dynamic range limited, causing convolution artifacts to appear 
        around the compact continuum peak; these artifacts are not real (see text).}
        \label{fig:cont}
\end{figure}

The resulting continuum maps from standard imaging (natural weighting,
Hogbom CLEAN algorithm) are shown in \figref{fig:cont}. IRAS~4A is a
resolved binary system \mbox{\citep{looney00,reipurth02,girart06}},
separated by 1\farcs8 where both components are relatively strong (NW
and SE). VLA~1623 is a triple system in the $\rho$~Ophiuchus star
forming region where one of the sources (source A) is the Class~0
object with a known Keplerian disk \citep{murillo13a,murillo13b}; the closest
companion is $\sim1$\arcsec\ away (source~B) and the second, which is significantly 
weaker at mm wavelengths, is $\sim10$\arcsec\ to the east (source~W). 

Toward IRAS~2A, \citet{codella14} reported two secondary continuum sources that are 
2\arcsec\ and 2.4\arcsec\ away, one along the blueshifted outflow and one
southwest of the main component. \citet{tobin15a} show that IRAS~2A is a binary 
on 0\farcs6 scales from high sensitivity VLA observations at mm and cm 
wavelengths, confirming earlier speculations based on the quadruple outflow 
\citep{jorgensen04}. The two secondary continuum sources (with 
$>$5$\sigma$ peak) found by Codella et al.\ at 2$''$ offset 
(sensitivity of 1.5~mJy/beam) are also not seen in our high-sensitivity 
PdBI data (1.6~mJy/beam). Our spatial resolution and 
sensitivity are not high enough to confirm the close companion found by 
\citeauthor{tobin15a} This unresolved companion is significantly weaker 
than the main source ($<3$\% at 1.3~mm,  i.e., 1.65$\sigma$); therefore, 
it will modify the amplitude in a well-behaved manner (a very low constant 
value).

The continuum observations of IRAS~4B are dynamic range limited, the 
peak flux is $128\sigma_\mathrm{RMS}$, and the limit for PdBI is at     
$\sim80\sigma_\mathrm{RMS}$ (depending on  \uv\ coverage, for example). This causes the 
low-flux compact deconvolution artifacts surrounding the continuum peak. This is 
not a problem since the analysis is conducted in the \uv-plane.

\begin{table}[ht]
\caption{Parameters for the continuum observations.}\label{tab:obs}
\centering
\begin{tabular}{l c c c c c}
\hline\hline          
        Source    & $\sigma_\mathrm{RMS}$ & Peak   &    Beam\\      
                          & (mJy/beam)                  &  ($\times\sigma_\mathrm{RMS}$)         & 
                          ($\arcsec$) \\      
\hline
   IRAS 2A    & 1.6 & 55 & $0.7\times0.9$ (63\degree) \\
   IRAS 4A    & 10  & 221 & $1.1\times1.3$ (90\degree) \\
   IRAS 4B    & 2   & 128 & $0.5\times0.6$ (37\degree) \\
   VLA 1623 A & 0.6 & 140 & $0.3\times0.5$ (92\degree) \\
\hline
\end{tabular}
\end{table}

%
\section{Models} \label{sec:method}

Images from radio interferometers combine the information obtained
from all baselines (scales) into one image. Thus, analyzing an image
of a small-scale structure whose emission constitutes only a fraction
of the total is not optimal. Visibilities provide a better starting
point for taking advantage of the full information of emission on
different size scales. This way the large-scale envelope can be
disentangled from emission on small scales. A large-scale, roughly spherical 
envelope model and possible companion source(s) are subtracted (and the data 
de-projected), then two different models are fitted to the residuals to study 
the small-scale structure in these five sources. The two fitted models are an 
intensity distribution consisting of a Gaussian plus a point source, and an 
analytical equation of the visibilities given by a passively heated thin disk. 
Simple power-law models to the visibilities without subtracting any large-scale 
envelope components are provided for comparison in Sect.~\ref{sect:envonly} for NGC~1333 IRAS~2A and in the Appendix for the rest of the sources.

\subsection{Envelope model}\label{sec:envmodel}

Class~0 sources are deeply embedded in their surrounding envelope
whose emission makes a non-negligible contribution to the total
flux. The envelope contribution to the flux falls off with decreasing scale
probed.

To analyze the smallest scales where the emission deviates from a
smooth symmetric structure, we first need to remove the larger scale
envelope contribution. This is done by simulating observations of model images 
of already published models for the envelopes of the studied sources and then 
subtracting them from the observations (see \figref{fig:example_prep} 
for an example and Appendix~\ref{app:steps} for all sources). The envelope 
models are taken to be spherically symmetric and constrained by a simultaneous 
fit to the SED and millimeter continuum radial profiles obtained with single-dish observations, probing scales of thousands of AU 
\citep{jorgensen02,kristensen12}. The most recent models include far-infrared 
measurements from {\it Herschel} to constrain the peak of the SED 
\citep{karska13}.  The density profile is a power law given as
\begin{equation}
        \rho_\mathrm{env}(r) = \rho_\mathrm{in} \left(\dfrac{r}{r_\mathrm{in}} \right)^{-p_\mathrm{env}}
,\end{equation}
where $\rho_\mathrm{in} = n_\mathrm{in}\, \mu_\mathrm{H_2}m_p $ and 
$\mu_\mathrm{H_2}\ = 2.8$ \citep[H$_2$, H, He, and 2\%\ metals;][ Appendix 
A.1]{kauffmann08}, $m_p$ the mass of a proton and $n_\mathrm{in}$ the number 
density at $r_\mathrm{in}$, the inner radius of the envelope.

Instead of interpolating the density and temperature to a finer grid 
from the published envelope models, a model image of the envelope was created 
from scratch for use in simulating the visibilities appropriate for the 
resolution of the interferometric observations. For a best-fit density 
distribution, we calculate the dust temperature self-consistently at each radius 
of the envelope for the observed luminosity using the dust continuum radiative 
transfer code TRANSPHERE \citep{dullemond02}. The dust density and temperature
structure is then input to the Monte Carlo radiative transfer code
RATRAN \citep{hogerheijde00} to produce the model image of the
continuum (assuming a gas-to-dust ratio of $\Delta_{g/d}=100$). The
opacities are taken from \citet{ossenkopf94}\footnote{Column 5  i.e.,\
  MRN distribution with thin ice mantles after $10^6$~years}.
This opacity at 203.4 and 219~GHz is 0.899~cm$^{2}$g$^{-1}$ and 2.57 at 335~GHz.

Because the envelope model is subtracted from the observational data,
the center position of the large-scale envelope needs to be
determined. This is assumed to be the continuum peak position in the
interferometric data or, if there are nearby companions, the center of
the large-scale emission of the system. This large-scale emission is
determined from the short \uv\ distances and is   
characterized by a Gaussian, the center of which is assumed to give
the center of mass of the large-scale envelope.

The resulting model image from the radiative transfer code is then
input, together with the shifted observed visibility data, to the
GILDAS routine \texttt{uv\_fmodel}. The routine simulates observations
of the model image by sampling the same \uv\ coverage as the
observations. This envelope model is then subtracted from the observed
visibilities. What is left is the contribution to the flux that cannot be attributed to such a spherical envelope alone, assuming that the 
envelope is optically thin.

\subsection{Gaussian intensity distribution}\label{sec:gausmodel}
As a first characterization of the visibilities after envelope subtraction, a 
Gaussian intensity distribution with an unresolved point source is fitted. The 
model corresponds to the emission from an embedded disk where the added 
unresolved point source is needed to reproduce the emission on the longest 
baselines. Most of the emission ($95\%$) of a Gaussian is emitted within 
$\pm1\sigma \sim 0.85\times \theta_\mathrm{FWHM}$. From the fit we can then get 
an estimate of the radius of the disk/compact structure, which is taken as 
$r_{c} = R = 0.42 \times \theta_\mathrm{FWHM}$. Expressing the FWHM 
($\theta_{FWHM}$) in arcseconds and the \uv\ distance ($r_{uv} = \sqrt{u^2 + 
v^2}$) in wavelengths, the function to fit a Gaussian intensity distribution 
with amplitude $F_{G}$ takes the form
\begin{eqnarray}
        F_\mathrm{re}(r_{uv}) = F_{ps} + F_{G}\exp\left[ -\dfrac{\left(\dfrac{\pi}{180}\dfrac{\pi\theta_\mathrm{FWHM}r_{uv}}{60^2}\right)^2}{4 \ln 2 }\right]
.\end{eqnarray}
The point source flux ($F_{ps}$) is simply a constant across all \uv\ distances 
(i.e.,\ Fourier transform of a Dirac delta function at the origin). This assumes 
that the structure is circularly symmetric and the center of mass is at the 
phase center, so the expected imaginary amplitude is zero for all baselines.

The gas mass of the structure at a distance $d$ is calculated as 
\begin{eqnarray}\label{eq:gaussmass}
        M = \dfrac{F_\nu\ d^2 }{\kappa_\nu B_\nu\left(T\right)} \Delta_\mathrm{g/d}
,\end{eqnarray}
where $F_\nu$ is the flux, $B_\nu(T)$ the Planck function at frequency
$\nu$ and temperature $T$, $\kappa_\nu$ the opacity at frequency $\nu$,
and $\Delta_\mathrm{g/d}$ the gas-to-dust-ratio. The mass is
calculated using the flux of the Gaussian and point source,  i.e.,\ $F_{ps} + 
F_{G}$, and assumes an average temperature of 30~K. This temperature could be 
lower, which would increase the mass \citep[e.g.,][]{jorgensen09,dunham14}

\subsection{Power-law disk}\label{sec:plawmodel}
The Gaussian model described above approximates the intensity distribution from 
a disk. Except for an estimate of the radius and the mass of such a disk, it 
does not fit any physical structure. It is possible to calculate the 
visibilities corresponding to an inclined circular disk with a power-law density 
and temperature profile \citep{berger07}. Thus, the fit will be sensitive to the 
physical conditions provided that such a disk is a good enough approximation of 
the physical structure. We note, however, that fitting observations at only one 
frequency is essentially a fit of $p+q$, i.e., the temperature and density 
together. To break the degeneracy, a temperature or density profile needs to 
be assumed or constrained by other means.

The model of the disk is at the phase center making the imaginary visibilities 
zero. A vertical disk structure is not included in the model. The radius of the 
disk, $r_c$, comes from the Gaussian fit previously described and the inner 
radius is set to 0.1~AU.

The temperature in the disk is given by
\begin{equation}
        T(r) = T_0 \left(\dfrac{r}{r_{T_0}}\right)^{-q},
\end{equation}
where $T_0$ is taken to be $1500$~K at $r_{T_0}=0.1$~AU. We  assume $q=0.5$ for 
all modeled disks. This is a reasonable value given both analytical studies and 
observations. A model of a radiative, passive circumstellar disk in hydrostatic 
equilibrium gives $q=0.5$ \citep[][extended model of 
\citealp{chiang97}]{chiang01}. Observations of more evolved disks (i.e.,\ T Tauri 
disks) show $0.4<q<0.75$ \citep{andrews05} confirming that $q=0.5$ is a 
reasonable value. This means that the 100~K radius is 
$r_{100~\mathrm{K}}=r_{T_0}(100/T_0)^{1/-q}$,  i.e.,\ 22.5~AU for the 
sources in this study. Such a large value of the water snowline is consistent 
with 2D radiative transfer models that explicitly include the accretional 
heating due to high mass accretion rates expected in Class 0 systems 
\citep{harsono15b}.
The lowest temperature allowed in the disk is $T_\mathrm{min}=10$~K. 
This is set by the external radiation field which heats the outer envelope to 
similar temperatures. The surrounding envelope at $r=300$~AU is roughly 
$T=20-50$~K, given by the envelope models of the sources (see 
Sect.~\ref{sec:envmodel} and Sect.~\ref{sec:powerlaw}). A lower 
disk temperature 
compared to the 
envelope at similar radii could be explained by the expected shielding of the 
disk mid-plane. However, $q$ has not been accurately 
determined in these systems, and is not necessarily constant with radius 
\citep[e.g.,][]{whitney03a}.

The surface (dust) density as a function of radius in the disk is given by 
\begin{eqnarray}
 \Sigma_\mathrm{disk}(r) =  \dfrac{\Sigma_0}{\Delta_\mathrm{g/d}} \left(\dfrac{r}{r_0}\right)^{-p} \qquad r < r_c
,\\
\Sigma_\mathrm{taper}(r) =  \dfrac{\Sigma_c}{\Delta_\mathrm{g/d}} \exp{\left[-\left(\dfrac{r}{r_c}\right)^{2-p}\right]} \qquad r \ge r_c
,\end{eqnarray}

where $r_0$ is the reference radius and $\Sigma_0$ (gas density) is expressed, 
and $\Sigma_c=\Sigma_\mathrm{disk}(r_c)\Delta_{g/d}$ beyond the disk 
critical radius $r_c$. The optical depth $\tau$ is given by the surface dust 
density multiplied with the opacity $\kappa_\nu$ and divided by the inclination 
$\cos i$,  i.e., $\tau (r) =  \Sigma(r)  \kappa_\nu/\cos i $. The visibility for a 
given \uv\ distance of such a disk can be calculated by integrating thin 
circular rings, each with a temperature and optical depth assuming blackbody 
emission \citep{berger07},

\begin{equation}
        V(r_{uv}) = \int \dfrac{2\pi{r}}{D^2} B_\nu(T(r))\ \cos i\ \left(1 - e^{-\tau(r)}\right)\ J_0(\dfrac{2\pi r_{uv}r}{D})\ dr
,\end{equation}
where $r_{uv}$ is the projected baseline expressed in wavelengths (cf.\ 
$\theta=1.22 \lambda/D$, where D is the baseline length), 
and all terms are expressed in cgs units. The taper can be 
seen as an approximation of the density of the disk-envelope interface, 
which could be due to many different factors, for example viscous spreading, 
non-spherical envelope contribution (i.e.,\ flattened inner envelope), and 
other relatively uncertain parameters. 
The free parameters in fitting this power-law disk are those for the 
surface density, $p$ and $\Sigma_0$. 

The gas mass of the disk (from radius $r_1$ to $r_2$) is calculated as
\begin{equation}\label{eq:plmass}
 M_\mathrm{disk}(r_\mathrm{in},r_\mathrm{disk}) = \dfrac{2\pi 
 r_0^p\Sigma_0}{(2-p)} 
 \left[r_\mathrm{in}^{2-p} - r_\mathrm{disk}^{2-p}\right].
\end{equation}
We note that the taper is excluded in the estimate. 

After fitting the power-law disk, it is possible to estimate the mid-plane 
density of such a passively heated structure, which is needed in order to discuss the disk 
chemistry and for comparison with the envelope model. We note that we do not fit the 
vertical structure.
For a passively heated vertically isothermal disk, the scale height is given as 
\begin{equation}
 h(r) = \sqrt{\dfrac{k_\mathrm{B} T(r) r^{3}}{G M_* m_\mathrm{p} \mu_\mathrm{H_2}} }
.\end{equation}
The density is then calculated for a central source of $M_{\star}=0.05~M_\odot$ 
for all sources except VLA~1623~A, where 0.2~\Msun\ is used 
\citep{murillo13b}. Thus, the mid-plane 
density and density at one scale height are given by
\begin{eqnarray}
\rho_\mathrm{midp}(r) = \dfrac{\Sigma_\mathrm{disk}(r)}{\sqrt{2 \pi} h(r) } \\
\rho_\mathrm{surf} = \rho_\mathrm{midp}(r)\ \mathrm{e}^{-\sfrac{1}{2}}
\end{eqnarray}
With these assumptions it is possible to look at how the density 
transitions to the envelope. Given no evidence of sharp edges in the 
visibility curve, the emitting surface should be smooth as well. The 
central stellar mass is not well constrained except for VLA~1623. 
However, increasing the stellar mass to\ 0.3~\Msun,\ for example, instead of the assumed 
0.05~\Msun\ for the relevant sources gives a higher midplane density by a factor 
of $\sqrt{6}\approx2.5$; this will not affect the conclusions drawn in this 
study.

%
\subsection{Application of models to data}

The analysis of the data is performed on the binned visibilities alone to 
avoid an additional uncertainty that the imaging methods can introduce (i.e., 
gridding and cleaning). Imaging was done to confirm phase shifts and companion 
subtraction. 
The size of the visibility bins was taken as the \uv\ distance corresponding to 
the individual antenna diameter.

Before fitting any structures to the observations the data are prepared in a few 
steps (illustrated for IRAS~4A-SE in \figref{fig:example_prep} and in 
Appendix~\ref{app:steps} for the rest of the sources): 

\begin{figure*}[htp]
        \centering
        \includegraphics[width=0.75\linewidth]{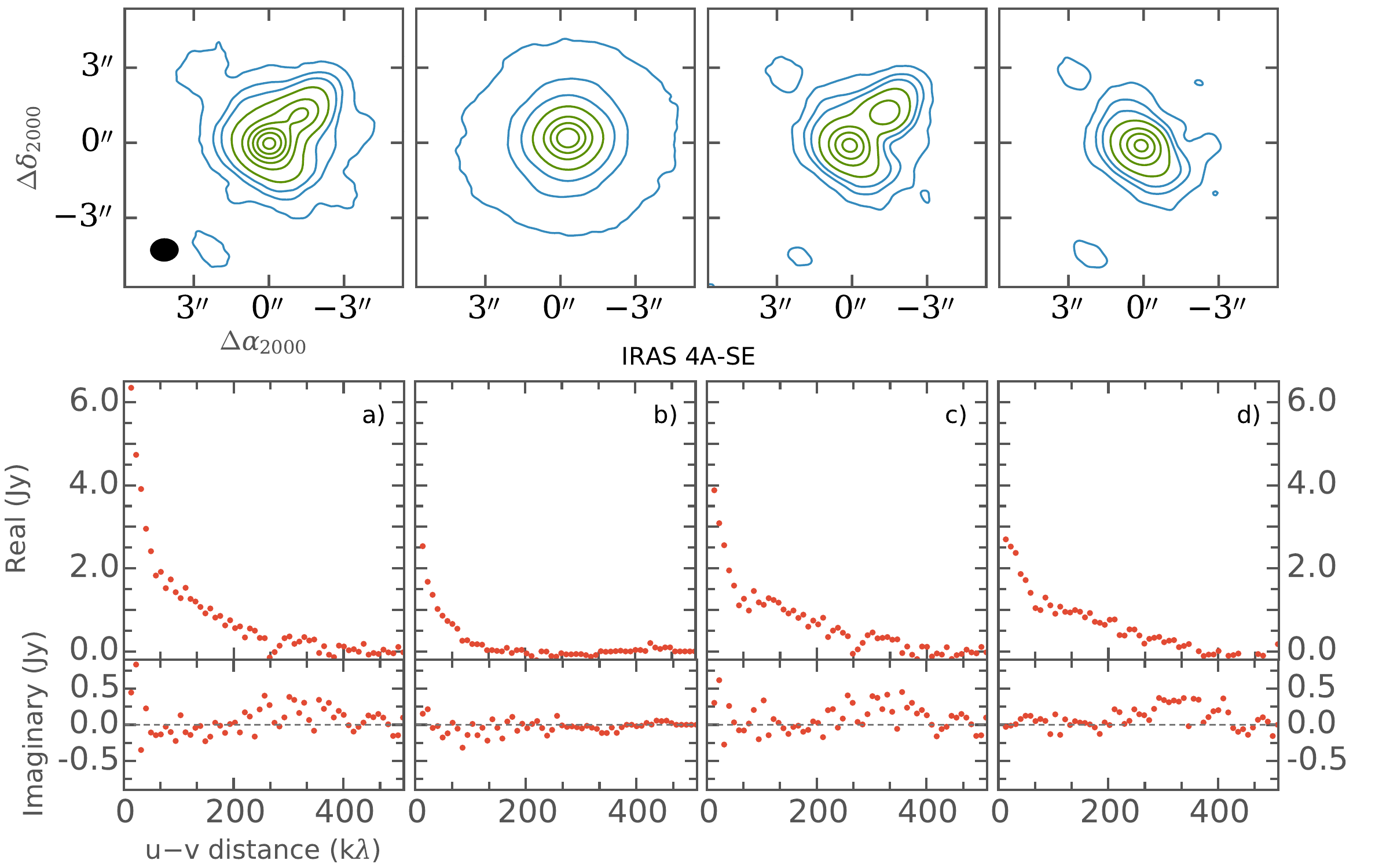}
        \caption{Images and visibilities of data preparation steps for IRAS~4A-SE. Top: imaged data; bottom: binned real and imaginary visibilities. From left to right: a) raw data, b) envelope model, c) raw data with envelope subtracted, d) raw data with envelope and companion subtracted. The contours start at 3$\sigma$, then in steps of 7$\sigma$ until 31$\sigma,$ and then in steps of 31$\sigma$, where $\sigma$=10~mJy (5~mJy for envelope model image).}
        \label{fig:example_prep}
\end{figure*}

\begin{enumerate}
\item A spherically symmetric model of the envelope is constructed. The model is 
subtracted from the observational visibilities (see Sect.~\ref{sec:envmodel} for 
details). 
\item A Gaussian and a point source are fitted to all components and those corresponding to companion(s) are subtracted from the visibilities. Visual inspection of the residual image confirms the success of the subtraction.
\item The phase center is shifted so that the continuum peak of the source of interest is at the center (relevant for companion systems).
\end{enumerate}
At this point, the visibility data of all sources consist primarily of emission 
not reproducible with the spherical envelope models and free of resolved close 
companion sources. IRAS~4A-SE/NW and VLA~1623~A both have nearby companions that 
are resolved and need to be subtracted. IRAS~4B has a weak companion at 
$\sim10$\arcsec\ directly east, but due to its weak nature and distance from the 
main 
source it should not contribute any real flux; however, it was subtracted before 
analysis. For IRAS~4A, where both sources have been analyzed and are relatively strong, 
there is a risk of left-over structure from the subtracted companion. However, 
the spatial scales involved in the fitting are smaller than the separation, and 
thus any structure left over should not affect the results.

The structure remaining after the envelope subtraction is assumed to be 
flattened and inclined with respect to the plane of the sky. Since the Gaussian 
and power-law disk models assume a face-on viewing angle when calculating the 
visibilities (circular Gaussian and face-on disk) the observational data need 
to be de-projected to be aligned with the plane of the sky. Since the 
minimization is done on the observations subtracted by the model, it is easier 
to de-project the data rather than the model. To de-project the model more  
calculation steps need to be performed to compare it with the observations 
because they need to be compared in the same \uv\ coordinates, which change 
when the visibilities are de-projected (see Appendix~\ref{app:orientation}). 
Assuming that the remaining compact axisymmetric structure is aligned 
approximately perpendicular to the large-scale outflow, the visibilities can be 
de-projected using the best estimates of the position angle (PA, measured east 
of north) and inclination. A face-on disk has an inclination of 0\degree\ and an 
edge-on disk 90\degree. Subsequently, the real amplitudes of the visibilities are 
analyzed, first by fitting an intensity distribution consisting of a circular 
Gaussian together with a point source. This provides an estimate of the radius 
and a mass of the compact structure. The estimate of the radius is then used in 
the fitting of the physical power-law disk model as the disk radius ($r_c$). The 
models are fitted to the data using the Levenberg-Marquardt algorithm through 
the SciPy.optimize module \citep{scipy}.

The de-projection is performed on the visibilities using the same approach as 
given in \citet{hughes07}, some details are given in 
Appendix~\ref{app:orientation}. If there are previous estimates of the PA and 
inclination (e.g.,\ from gas tracers showing rotation) these values are used. Otherwise 
it is assumed that 
$\mathrm{PA}_\mathrm{disk}=\mathrm{PA}_\mathrm{outflow}+90\deg$. Furthermore, 
for all sources except VLA~1623~A and IRAS~2A, there are no reliable estimates 
for the inclination angle, which is then assumed to be $60\deg$. However, 
changes in the inclination of 15\degree\ are run to assess the effects 
of inclination (45\degree and 75\degree). Current best estimates 
of the PA and inclination from literature, where applicable, are listed in 
\tabref{tab:painc}. We note that inclinations close to edge-on,  i.e.,\ 90\degree\ 
are not easily constrained with this method ($\lim\limits_{i \to 90} 
\cos^{-1}i\to\infty$).

\begin{table}[ht]
\caption{Adopted position angle and inclination of compact component.}\label{tab:painc}
\centering
\begin{tabular}{l c c c}
\hline\hline          
Source  & PA$_\mathrm{disk}$ & inc & Ref. \\      
\hline                       
   IRAS 2A              &  $110\deg$ &    $58.5\deg\pm15$    & 1\\
   IRAS 4A-SE/NW&        $100\deg$               &    $60\deg\pm15$  & 2 \\ 
   IRAS 4B              &   $90\deg$             &   $60\deg\pm15$    & 2 \\
   VLA 1623     &  $35\deg\pm5$  & $55\deg\pm5$ & 3 \\

\hline
\end{tabular}
\linebreak
\tablefoot{1. \citet{tobin15a} VLA 8 and\ 9mm A, AB configuration data.
2. By eye estimate from CO (2-1) outflow $+90\deg$ \citet{jorgensen07} and 
\citet{yildiz12}.
3. Kinematic line analysis \citet{murillo13b} 
}

\end{table}

%
\section{Results}

\subsection{Gaussian intensity distribution}\label{sec:gaufit}
\begin{figure*}[ht]
        \centering
        \includegraphics[width=0.7\linewidth]{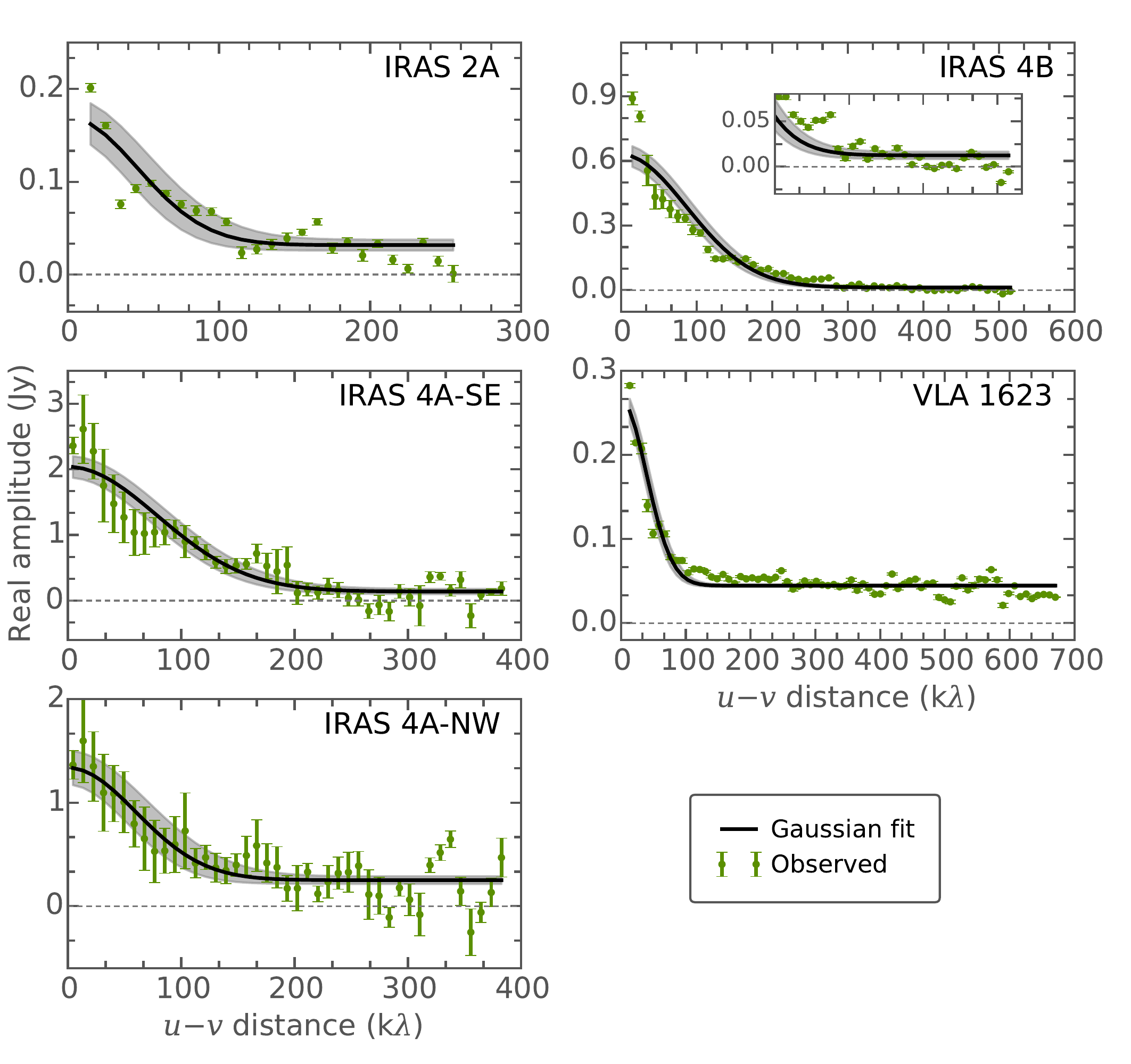}
        \caption{Best-fit Gaussian model with added point source where applicable. 
        The green points are the envelope subtracted and de-projected data with the 
        variance from the binning as uncertainty, while the black line and gray area 
        show the best fit and the 1$\sigma$\ region (changing the fitted 
        parameters by $\pm$1$\sigma$). }
        \label{fig:gauss_fit}
\end{figure*}

The results of fitting a Gaussian intensity distribution plus point
source are shown in Figure~\ref{fig:gauss_fit}. The parameters of the
best fits are presented in \tabref{tab:gauss_pars}. The radius is
calculated from 0.42 $\times \theta_{\rm FWHM}$.

In fitting a point source it is assumed that the most compact
structure is not resolved. The data of IRAS~2A and IRAS~4A-SE/-NW are
 represented well by the fit function. However, for these sources
information on the longest baselines, as available for VLA~1623~A and
IRAS~4B, is missing. Furthermore, the data for IRAS~4A-NW/-SE have a 
lower signal-to-noise ratio. IRAS~4B has a jump in real amplitude around
280~\klam; the reason for this is unclear. Toward the longest
baselines, the IRAS~4B amplitude seems to go below zero, signifying that 
perhaps the compact structure has a sharp edge or that the center of 
mass is slightly offset from the phase center; longer baseline data could shed 
light on this. 
For VLA~1623~A, the non-zero values out to long 
\uv\ distances show that a compact structure is probably present. The Gaussian 
plus point source fit is worse in the range $70\sim250$~\klam\ for VLA~1623~A.

\begin{table*}[ht]
\caption{Parameters from Gaussian fits, with uncertainty in the values reflecting changes in inclination by $15\deg$.}\label{tab:gauss_pars}
\centering
\begin{tabular}{l c c c c c c c}
\hline\hline          
        Source          & $F_G$ & \multicolumn{2}{c}{$\theta_\mathrm{FWHM}$} &  $F_\mathrm{PS}$ & $R$ (=$r_c$)  & $M_\mathrm{gas}$ \\      
                                & [mJy]  & [$\arcsec$] & [AU] & [mJy] & [AU]  & [$\times10^{-3} M_\odot$] \\      
\hline
   IRAS 2A              & $140$\err{40}{20}     & $1.7$\err{0.6}{0.2}   & $396$\err{150}{50}& $30\pm10$                   & $167$\err{63}{21} & $155$\err{46}{27}  \\
   IRAS 4B              & $600$\err{60}{40} & $0.9\pm0.1$                       & $210\pm15$              & $\leq10$                              & $88\pm6$                 & $556$\err{55}{46}  \\
   IRAS 4A-SE   & $1900\pm100$          & $0.98$\err{0.32}{0.26}& $230$\err{75}{60} & $100$\err{100}{60}    & $96$\err{32}{25}      & $262$\err{26}{21}      \\
   VLA 1623     & $220\pm16$            & $2.0$\err{0.9}{0.4}   & $237$\err{105}{46}& $40\pm2$                                & $100$\err{44}{20} & $54\pm4$                   \\
   IRAS 4A-NW   & $1100\pm160$          & $1.3$\err{0.2}{0.1}   & $303$\err{92}{41} & $300$\err{40}{100}    & $127\pm18$            & $183$\err{26}{34}      \\
\hline
\end{tabular}
\tablefoot{
$R$ is taken to be $0.42\times\theta_\mathrm{FWHM}$. The masses are derived using the $F_{G}+F_{ps}$ flux, and assuming $T_\mathrm{avg}=30$~K.
}
\end{table*}

\subsection{Power-law disk}
\label{sec:powerlaw}

\begin{figure*}[htp]
        \centering
        \includegraphics[width=0.7\linewidth]{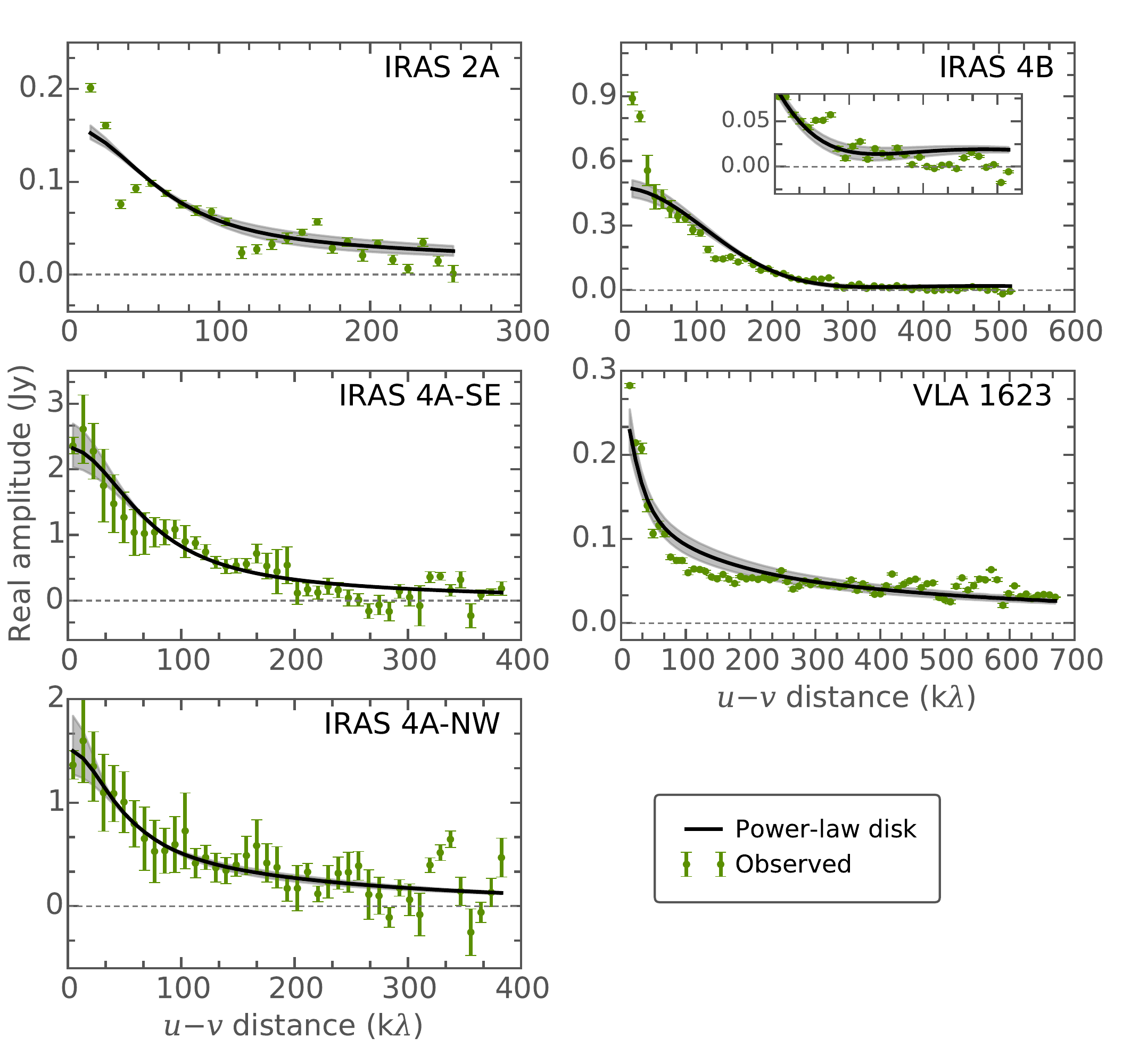}
        \caption{
        Resulting best-fit power-law disk for the sources. The green points are the 
        envelope subtracted and de-projected data with the variance from the binning 
        as uncertainty, while the black line and gray area show the best fit and the 
        1$\sigma$\ region (changing the fitted parameters 
                by $\pm$1$\sigma$).}
        \label{fig:plaw_fit}
\end{figure*}

The resulting best-fit power-law disk model to the envelope subtracted
and de-projected real visibility amplitude curves for each source are
presented in Figure~\ref{fig:plaw_fit}. The best-fit parameters 
are summarized in \tabref{tab:plawinfo}. For $r_c$, the best-fit value from the 
Gaussian fit is taken. We note  that the power-law disk
fits the temperature multiplied with the density,  i.e.,\ $p+q$, with $q$
fixed to 0.5 before fitting $\Sigma$ and $p$. The density fall-off in 
the taper is significantly steeper and this is reflected in the increase in 
intensity in the visibilities. The values for the density power-law index, $p$ 
is similar for all sources ($\sim1$) except IRAS~4B. The gas surface density 
varies between 2.4 and 51~g~cm$^{-2}$ at $r_0= 50$ AU.

IRAS~2A and IRAS~4A-NW/-SE are well represented with the disk model, with 
similar constraints on the disk parameters. The fit is worse in the range 
$70\sim200$~\klam\ for VLA~1623~A and IRAS~4B. This is partly because of the 
addition of a taper, whose radial density decrease depends on the disk density 
(i.e., $2-p$). While not ideal, it should not have a significant effect 
on the estimated disk parameters and the subsequent discussion on the derived 
physical conditions.

\begin{figure}[htp]
        \centering
        \includegraphics[width=\linewidth]{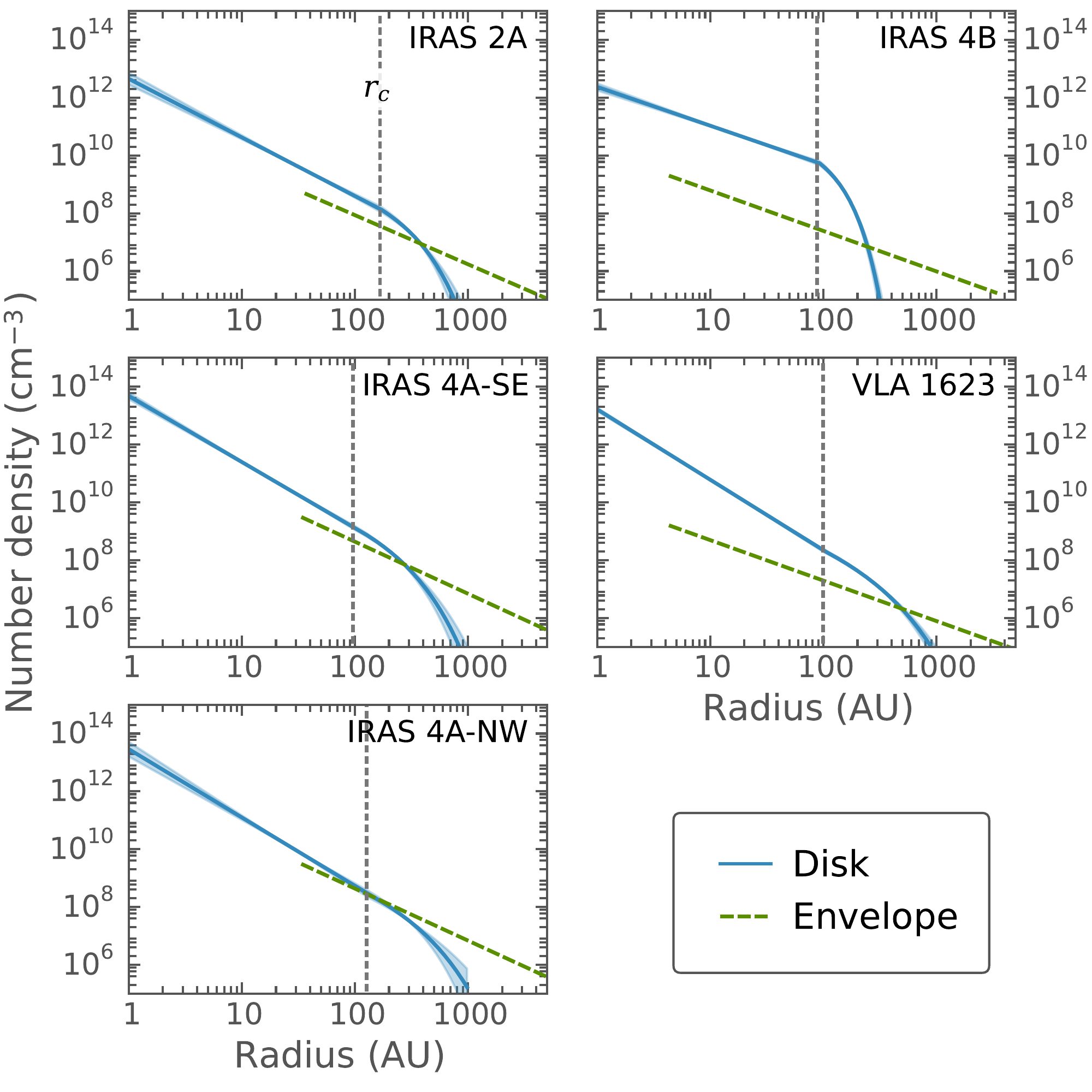}
        \caption{Midplane gas volume density of the disk (blue) and the envelope (dashed green) as a function of radius. The blue shaded area shows the uncertainty in the number density varying parameters $\pm1\sigma$. The volume density at the disk surface is a factor of $e^{-\sfrac{1}{2}}\sim0.6$ lower.}
        \label{fig:plaw_rho}
\end{figure}

\begin{figure}[htp]
        \centering
        \includegraphics[width=\linewidth]{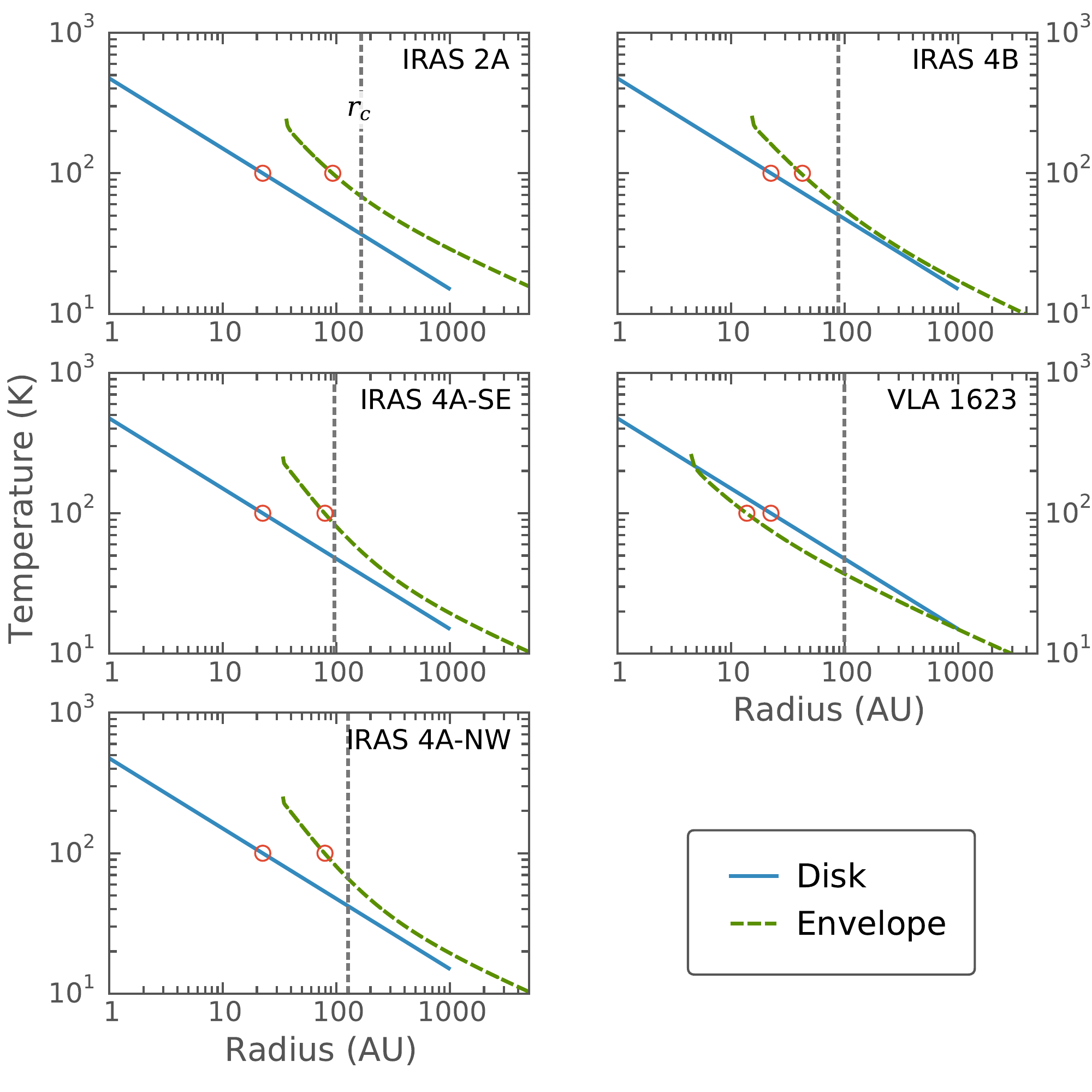}
        \caption{ Midplane temperature of the disk (blue) and the envelope 
        (dashed green). The red circles mark where $T=100$~K and the vertical dashed 
        line indicates the radius of the compact component ($r_c$).}
        \label{fig:plaw_temp}
\end{figure}

The corresponding best-fit density and temperature as a function of radius for 
each source are shown in Figures~\ref{fig:plaw_rho} and \ref{fig:plaw_temp}. The 
largest jump in density between the disk and envelope is seen in IRAS~4B where 
the interface between them is not smooth. The disk derived for IRAS~4B is small 
and has a shallow density profile. As noted, the vertical disk structure is not 
fitted in this study, so the mid-plane disk density connects with the envelope 
through the 
taper. The continuum $\tau=1$ radius at 1.5~mm lies within $R\sim10$~AU for
all sources and will not affect our results for the larger scales. 

The midplane temperature profile does not align well with the envelope
for IRAS~2A (\figref{fig:plaw_temp}), but does merge smoothly for the
other sources. Disk shadowing will affect both temperature and density, for example,\ 
lowering the temperature just behind the disk, which could explain the jump in 
temperature and density in some of the disk-envelope interfaces shown here
\citep[e.g.,][]{murillo15}. It is  important to note again here that the 
midplane gas volume density is derived using a combination of fitted and fixed 
parameters and it should be kept in mind when interpreting the results. The 
midplane density depends on the scale height, which is a function 
of the temperature profile (which is fixed, see Sect.~\ref{sec:plawmodel}).

\begin{table*}[ht]
\caption{Parameters from power-law disk fitting; the ranges reflect changes in 
the assumed $r_\mathrm{c}$, inclination, and PA from the fitting of a Gaussian 
and point source (see  section \ref{sec:gaufit}).}\label{tab:plawinfo}
\centering
\begin{tabular}{l c c c c c}
\hline\hline          
        Source          & $p$           & $\Sigma_0$    & $M_\mathrm{disk}$ & 
        min($T>100$~K)\\     
                                &                       &   [g cm$^{-2}$]       &  [$\times10^{-3} M_\odot$]                       & \%   \\       
\hline
   IRAS 2A              & $0.8$\err{0.4}{0.2}   & $6.7$\err{1.2}{0.4}   & $42$\err{9}{5}          & $9$\err{6}{3} \\
   IRAS 4B              & $\leq0.1$\err{1.1}{}  & $51$\err{15}{5}       & $140$\err{105}{15}      &  $7.4$\err{26}{0.3} \\
   IRAS 4A-SE   & $1.0\pm0.1$                   & $26\pm3$      & $90$\err{34}{20}         &  $25$\err{12}{6}      \\
   VLA 1623     & $1.2\pm0.2$                   & $2.4\pm0.5$   & $9\pm1$                         & $30\pm3$      \\
   IRAS 4A-NW   & $1.1$\err{0.1}{0.2}   & $11.3$\err{2}{1.2}    & $52$\err{11}{4}         &  $22$\err{1}{2}       \\
\hline
\end{tabular}
\tablefoot{
The temperature power-law dependence is fixed to $q=0.5$, as described in the 
text. $\Sigma_0$ is the gas density.
}
\end{table*}

%

\section{Discussion}\label{sec:discussion}

In this study, the de-projected continuum interferometric visibilities of five 
deeply embedded low-mass protostars are fitted with two different models, 
a Gaussian disk intensity distribution and a parametrized analytical expression 
of a vertically isothermal thin disk model. Before fitting the disk models, the 
current best model of the surrounding large-scale envelope and possible 
companion sources were subtracted.

\subsection{Method uncertainties}
There are various uncertainties related to the analysis which should be 
re-iterated before an in-depth discussion of the results.

The analyzed envelope subtracted residual visibilities depend on the 
envelope model used. The most recent models by \citet{kristensen12} and 
\citet{jorgensen02} are used to minimize this source of uncertainty. The models 
use both the observed SED and radial profile of the millimeter emission to 
constrain the envelope density and temperature profile.

The disk mass estimate, Equation~(\ref{eq:gaussmass}), assumes an 
average temperature of the disk, which is the same for all sources. For simple 
cases this should be a good enough approximation, and facilitates easy 
comparison with results from other studies of the same or similar sources where 
the same method was used.

The assumed stellar mass and temperature profile of 
the power-law disk will affect the derived midplane density, 
and to some extent the surface density profile (i.e.,\ $p$) given the 
degeneracies of calculating the visibilities (i.e., $p+q$ is fitted). The effect 
of the stellar mass on the derived midplane density is small. The assumed 
temperature profile also affects the midplane density, in the opposite way from 
mass. Increasing both the stellar mass and disk temperature by a factor of two 
will not have any effect on the derived midplane density. However, the assumed 
temperature profile has an impact on the amount of material above 100~K and thus 
the derived molecular abundances. As discussed in Sect.~\ref{sec:plawmodel} we 
have assumed a temperature profile that is reasonable and in agreement with 
other studies. Future higher sensitivity and higher resolution observations of 
molecular lines sensitive to the temperature structure can potentially improve 
the constraints.

\subsection{Disk radii and masses}
The derived range of radii of the disk-like structures, $90-170$~AU,
are plausible and are marginally resolved in these data sets. The disk 
diameter ($2{\times}r_c$) corresponds to the scale where there is a sharp
increase in flux.

For the one source, VLA~1623~A, for which the radius of the Keplerian
disk has been determined from kinematics of C$^{18}$O line emission, a
value of $R=150$~AU (rotationally supported to $R=180$~AU) has been
found by \citet{murillo13b}. The radius derived in this study is
$r_c=100^{+44}_{-20}$ AU, which is somewhat smaller but in agreement
within the uncertainties. In more mature disks, the mm dust continuum 
has been found to be more centrally concentrated than the gas due to radial 
drift \citep[e.g.,][]{andrews12}. This could be happening in VLA~1623 
even at this early stage.

The inferred masses vary between the sources, and are
different depending on the method. The mass estimates using the Gaussian 
intensity distribution are generally higher ($54-556\times10^{-3}$~\Msun) than 
for the power-law disks ($9-140\times10^{-3}$~\Msun). While the mass 
estimates for the power-law disk use the fitted disk 
parameters put into Equation~(\ref{eq:plmass}), the mass from the Gaussian 
intensity distribution is simply the Gaussian with the point source flux added 
put into Equation~(\ref{eq:gaussmass}).

\citet{jorgensen05b} fitted a Gaussian intensity distribution together
with a point source to 350~GHz SMA continuum observations of IRAS~2A
covering baselines between 18 and 164~\klam. They found a radius of
this compact emission of $100 - 200$~AU and a mass of 0.1~\Msun\ (for
$T_\mathrm{avg}=30$~K and flux of \uv\ dist $>50$~\klam). Our disk 
radius and mass derived from the Gaussian intensity distribution agree with 
this estimate.

In IRAS~4B the relatively high mass derived with the power-law disk shows a small
disk with a flat density distribution and a high surface
density. \citet{choi11} argue from the extent of the 1.3~cm continuum
emission toward IRAS~4B that the radius of the compact disk-like 
structure is $R=25$~AU, roughly 1/3 of the disk size derived here. Larger 
grains, similar to those responsible for the emission at these 
wavelengths, are affected by radial drift to a higher extent 
\citep[e.g.,][]{birnstiel10}, and the 
cm emission is thus expected to be more compact, in agreement with these results. 
However, \citet{choi10} derived a radius of at least 220~AU from kinematic 
analysis of ammonia emission at 24~GHz (1.3~cm).

\citet{jorgensen09} estimated the mass of the compact components toward
IRAS~2A ($56\times10^{-3}$~\Msun), IRAS~4A-SE
($460\times10^{-3}$~\Msun) and IRAS~4B
($240\times10^{-3}$~\Msun). This was done by using a combination of the
interferometric flux on 50~\klam\ scales and single-dish flux at
850~$\mu$m. The masses derived from the power-law disk in this
study agree for IRAS~2A and within a factor of 2 for IRAS~4B, but the 
disk mass derived for IRAS~4A-SE is five times lower. 
The gas masses derived from the Gaussian intensity distribution are about a factor of 2-3 
higher than that of \citeauthor{jorgensen09} for IRAS~2A and 
IRAS~4B, but for IRAS~4A-SE it is roughly half. The 
generally higher masses derived using the Gaussian intensity distribution comes 
from the fact that \citeauthor{jorgensen09} uses the flux at 50~\klam\ to derive 
the disk mass and a steeper power-law envelope profile; \figref{fig:gauss_fit} 
shows that this flux is still not the total flux of the compact component left 
after subtracting the spherically symmetric envelope contribution.

The fitted density power laws show that there is a significant amount of 
material at large radii. This is important for fragmentation. Numerical studies 
have shown that if the ratio of disk to stellar mass ($M_\mathrm{disk}/M_*$) 
goes above $\sim0.25$ the disk can be prone to gravitational instabilities  
\citep[e.g.,][]{lodato04,dong15}. It has been suggested that any gravitational 
instabilities may induce the formation of multiple star systems  
\citep[e.g.,][]{kratter10}. These instabilities can only form in disks 
more massive than in the Class~II stage (i.e.,\ higher $M_{disk}/M_*$), thus more 
likely in the earlier protostellar stages when the stellar mass is still low. 
Given that about half of all solar analogs are found in multiple systems 
\citep{raghavan10}, determining the physical conditions in the early stages of 
disk formation is of great importance for the understanding of the formation of 
multiple star and planetary systems. 
For VLA~1623, where the stellar mass was determined previously, the 
$M_\mathrm{disk}/M_*$ ratio is $0.009/0.20=0.045$, indicating a gravitationally 
stable disk. All the other sources have values higher than 
$M_\mathrm{disk}/M_*=0.25$, assuming $M_*=0.05$. With $M_*=0.2$ as for 
VLA~1623, the disks have ratios close to or higher than $0.25$, indicating that 
the disks may be gravitationally unstable even in this scenario.

For all sources, the envelope is relatively well characterized on
large scales, since both radial profiles and the SED were used to model the 
large-scale envelope. It is not clear whether the added taper, which is frequently 
used in similar studies for more evolved disks (i.e.,\ Class~II disks) 
\citep{andrews08}, is a good representation of the disk-to-envelope interface 
for these sources. However, it is also important to characterize this interface region that is not fitted by spherical envelope models. For some
of the sources in the study there are no observations with longer
baseline coverage, thus adding degeneracy to the fitting

It is clear from the various data sets used here that it is important
to cover longer \uv\ distances to constrain the disk. Of all the observations, only the ALMA data on VLA~1623~A cover baselines that fully resolve the disk-like structure, the observations of the other sources should still be sensitive to the structure but to a less extent. For more robust models of the radial structure, higher resolution observations are needed.

\subsection{Envelope-only model}\label{sect:envonly}
Recently, \citet{maury14} fitted the 1.3~mm continuum visibility amplitudes of 
IRAS~2A obtained with PdBI (baselines 14 to 557~\klam, 8\arcsec-0{\farcs}35) 
with a power-law intensity distribution, representing a single spherical 
envelope. The intensity as a function of \uv\ distance for an envelope with 
$T\propto{r^{-q}}$ and
$\rho\propto{r^{-p}}$ is $V\propto{b^a}$, where $a=(p+q-3)$. To
compare to these results, such a power-law envelope was fitted to the
visibilities of IRAS~2A in this study. It is important to note that the
two companions that \citeauthor{maury14} subtract prior to
envelope-fitting do not exist in our data (see
Sect.~\ref{sec:sample} for references), so we do not subtract any
companion sources for this analysis. 

\begin{figure}[htp]
        \centering
        \includegraphics[width=0.9\linewidth]{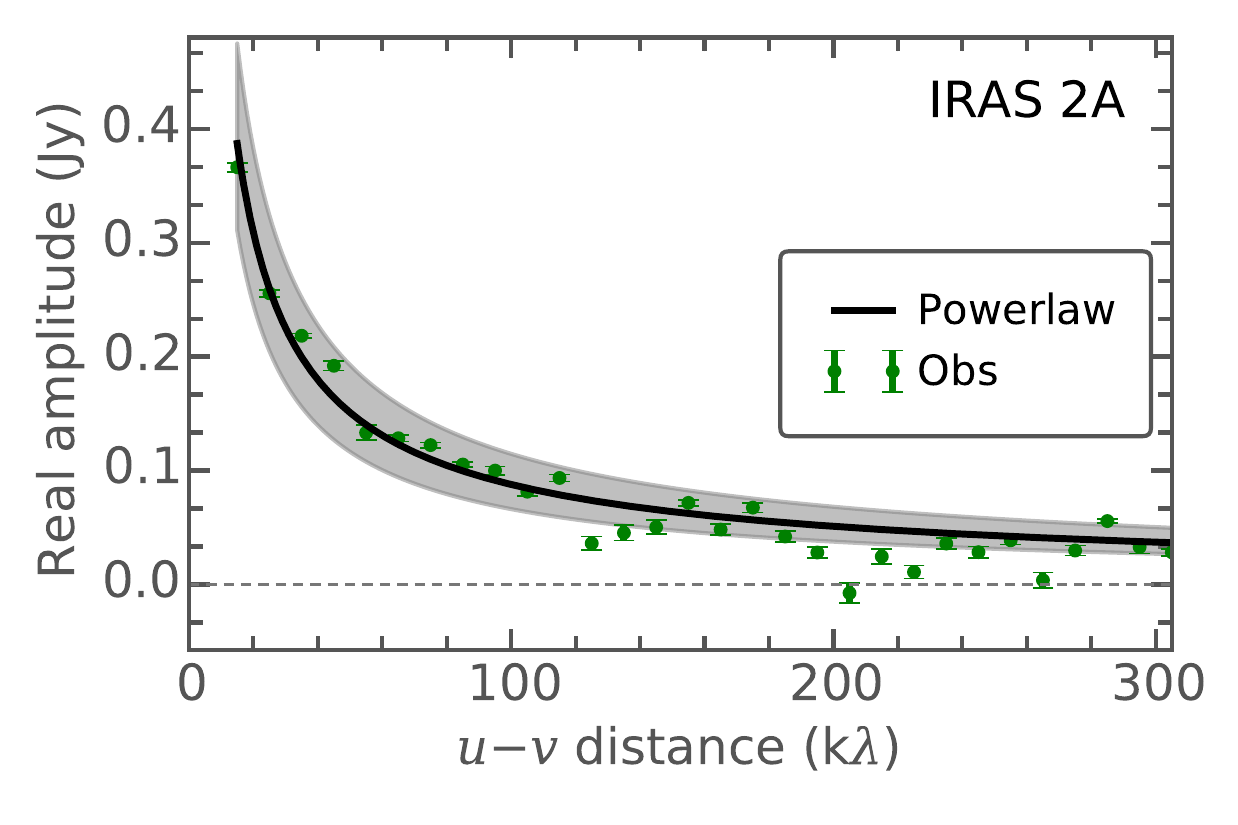}
        \caption{IRAS 2A envelope power-law model, $V=C b^a$, applied to the IRAS 2A data without subtraction of our envelope model. Here $a=-0.78\pm0.03$,  $C=3.2\pm0.4$, and $b$ is in units of k$\lambda$.}
        \label{fig:env_plaw_i2a}
\end{figure}
The resulting power-law envelope fit is shown in
\figref{fig:env_plaw_i2a}, and the fit coefficients are $a=-0.78\pm0.03$ 
($V=C{b^a}$, where $C=3.2\pm0.4$ if $b$ is in
k$\lambda$). This implies $p+q=2.22$, and for comparison,
\citeauthor{maury14} presented $a=-0.45\pm0.05$,  i.e.,\ $p+q=2.55$. The
envelope-only power-law fits for the rest of the sources in this study 
are presented in Appendix~\ref{fig:env_fit_all} and \tabref{tab:env_only}.  

\begin{figure}[htp]
        \centering
        \includegraphics[width=0.86\linewidth]{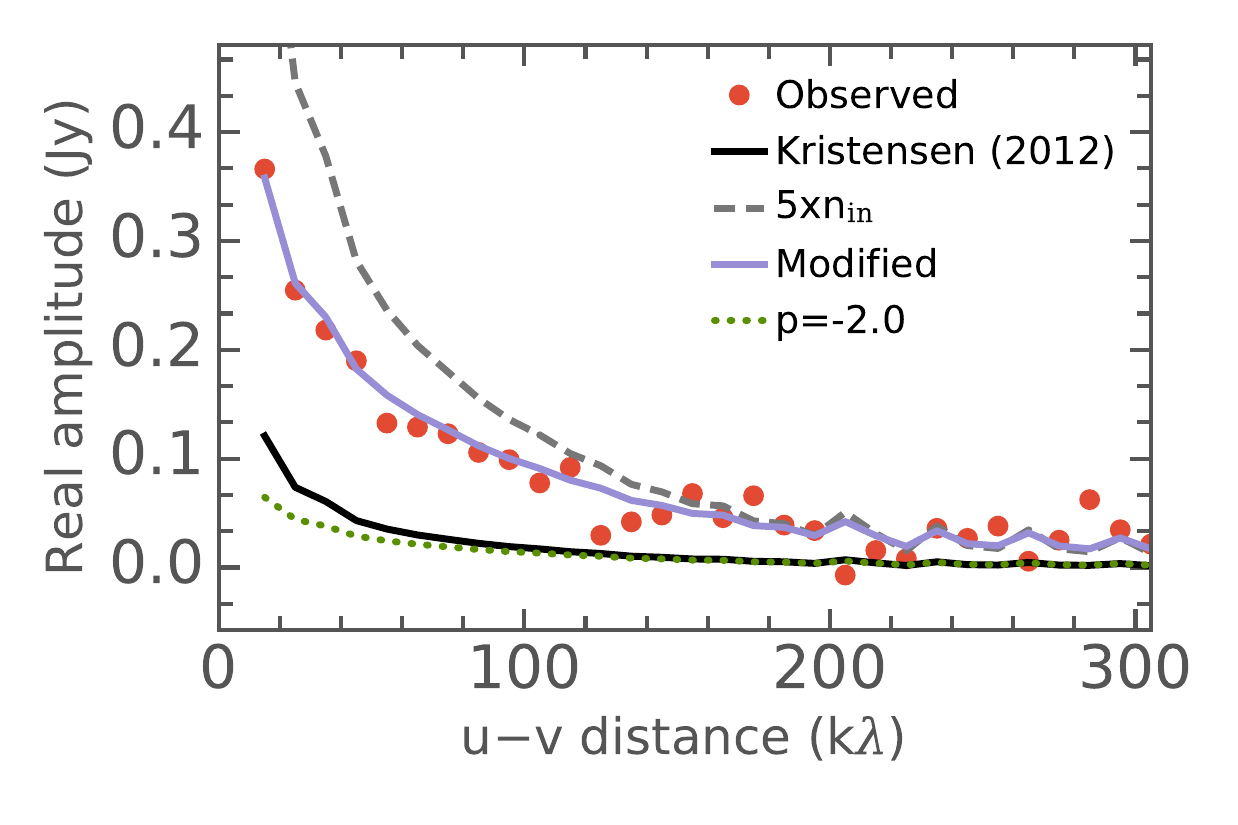}
        \caption{
        Modified envelope-only models from \citet{kristensen12} in comparison to the 
        observed interferometric visibilities of IRAS~2A. The non-modified  
        model of \citeauthor{kristensen12} is shown as a black line;  the by eye 
        best-fitting modified model, with $5\times$ higher number density 
        ($n_\mathrm{in}$) at the 
        inner radius and $p=-2.0$ is shown in purple.}
        \label{fig:envvis_models_i2a}
\end{figure}

While the modified power-law envelope model can reproduce the interferometric visibilities, it does not mean that the entire envelope is reproduced with this model alone. Modifying the envelope model by \citeauthor{kristensen12} to fit the visibility amplitudes gives the results shown in \figref{fig:envvis_models_i2a}. The model that best reproduces the interferometric visibility amplitude is where the volume number density is increased by a factor of 5 from $4.9\times10^{8}$ to $25\times10^{8}$~cm$^{-3}$ and the power law steepened from $p=-1.7$ to $-2.0$. This essentially represents moving more of the mass inward to accommodate the flux on longer baselines,  i.e., smaller scales. The wiggles at large baselines are due to the sharp edge introduced at the inner radius.

With this modified model the analytical prediction of the temperature 
slope ($q$) from the single powerlaw  envelope fit of \citeauthor{maury14} would 
be $2.55 - 2.0=0.55$. The resulting temperature profile from TRANSPHERE is 
roughly approximated by $q=0.5$. This 
shows that the modified model of the envelope is consistent with that of 
\citeauthor{maury14} on these aspects.

The observed flux at 219~GHz given by \citet{maury14} \citep[scaled 
from][]{motte01}, is 0.86~Jy. The total flux at 219~GHz after convolving with the
11\arcsec\ beam is 1.02~Jy for the \citeauthor{kristensen12} model and 
1.25~Jy for the modified interferometric model. Given a typical 
absolute flux uncertainty in single-dish observations of about 20\%, 
the modified model is higher by $45\%$, while the \citeauthor{kristensen12} model 
is higher by $19\%$.  The modified model is higher by a significant amount, although  not enough to discard it. Thus the modified single power-law envelope can 
reproduce the interferometric visibilities and almost the total flux. 

\begin{figure}[htp]
        \centering
        \includegraphics[width=0.9\linewidth]{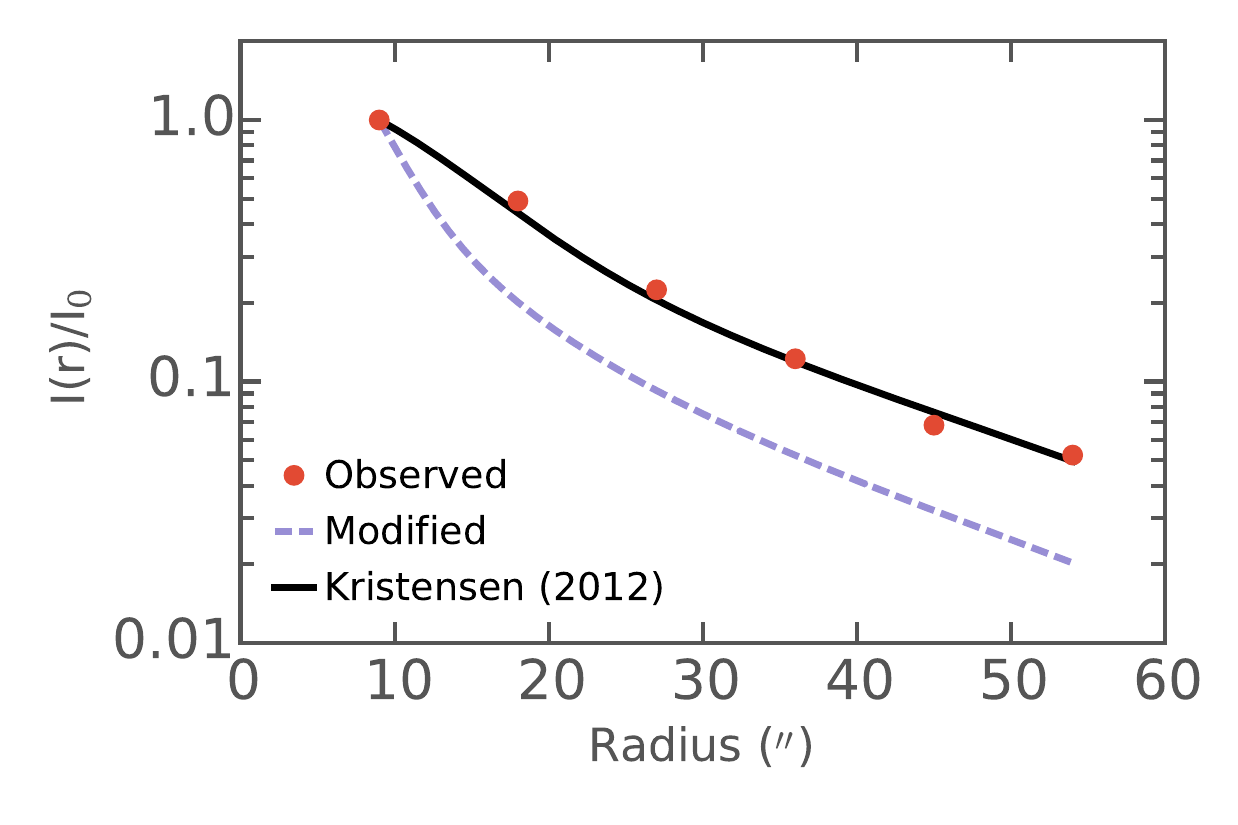}
        \caption{
        Radial profile at 850~$\mu$m for IRAS 2A (red dots, FWHM$_\mathrm{beam}=$ 
        19\farcs5) with the power-law envelope model of 
        \citet{kristensen12} (black line), and the modified 
        \citeauthor{kristensen12} model (purple dashes, $5\times 
        n_\mathrm{in}$ and $p=-2.0$).}
        \label{fig:rprofile_i2a}
\end{figure}
The envelope models of \citeauthor{kristensen12} and \citet{jorgensen02} are not 
only constrained by the total flux (SED), but also by the radial 
profile of the large-scale emission. The radial profiles of the two 
envelope models are shown in \figref{fig:rprofile_i2a} together with 
observations at 850~$\mu$m for IRAS~2A. 

While the model of \citeauthor{kristensen12} fits the radial profile, on scales 
larger than about 20\arcsec\ the modified model is at least 5 times lower than 
the observations. This is not surprising since much of the mass has been moved 
inward to fit the compact component of the emission. Thus, the single power-law 
envelope cannot reproduce the interferometric visibilities and the large-scale 
emission at the same time. Additionally, the morphology of IRAS~2A shows 
non-circular symmetric structures already at scales of 2\arcsec\ in 1.3mm 
observations \citep{tobin15b}, demonstrating that simply modifying the inner 
radial density distribution of the power-law envelope is not enough. Thus, a 
compact structure, possibly disk-like, together with an extended envelope 
provides the best fit to the total observed emission.

\subsection{Mass above 100 K}
\begin{figure}[htp]
        \centering
        \includegraphics[width=\linewidth]{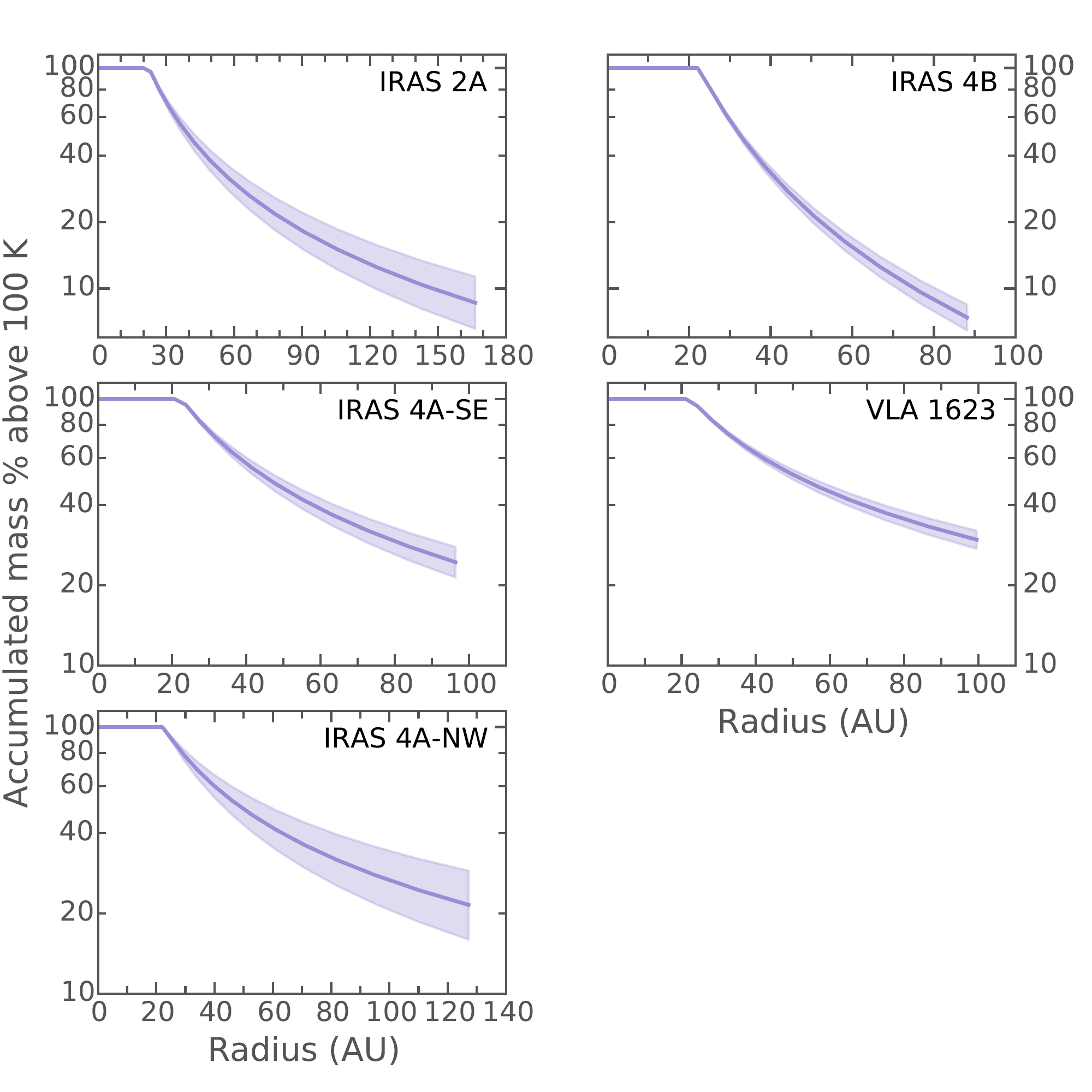}
        \caption{ Percentage of cumulative mass that is above 100~K as a 
        function of radius for the sources studied. The shaded area reflects the 
        mass introduced by  the uncertainty in the fitted disk parameters 
        with $\pm1\sigma$  ( i.e., 
        $\Sigma, p$).}
        \label{fig:plaw_tmass}
\end{figure}

The percentage of the material that is above 100~K for each radius of
the disk is computed based on the best-fit power-law disk
model. Figure~\ref{fig:plaw_tmass} presents the mass enclosed for
each radius compared with the mass where $T\ge 100$~K.
The radius where $T=100$~K is the same for all sources, given the
assumption that $q=0.5$ and a fixed $r_{T_0}$ (effectively fixed 
$L_*$). The different sizes and densities of the disk then change the percentage 
of the warm material at a certain radius. The $\sim20\%$ limit lies between 
$50\sim150$~AU for the sources except for VLA~1623~A, which never goes below 
30\%. In Table~\ref{tab:plawinfo} the percentage at the disk radius ($r_c$) is
listed for the studied sources. The percentage can be relatively
large, up to 30\%\ (VLA~1623~A) and as low as 7\%\ (IRAS~4A-SE). Changing the 
temperature structure such that the 100~K radius is at 5~AU but 
leaving everything else the same lowers the percentage  to 9\% for 
VLA~1623~A and to 0.5\% for IRAS~4A-SE. 

\subsection{Water abundance}

As mentioned in the introduction, the assumed physical structure
affects the abundance estimates. The amount of material with dust
temperatures above $\sim 100$ K is particularly important since this
is the temperature above which water ice sublimates and any
complex molecules locked into the ice are released \citep{fraser01}  (the specific temperature depends on density, see \citealt{harsono15b}).

\begin{figure}[htp]
        \centering
        \includegraphics[width=0.9\linewidth]{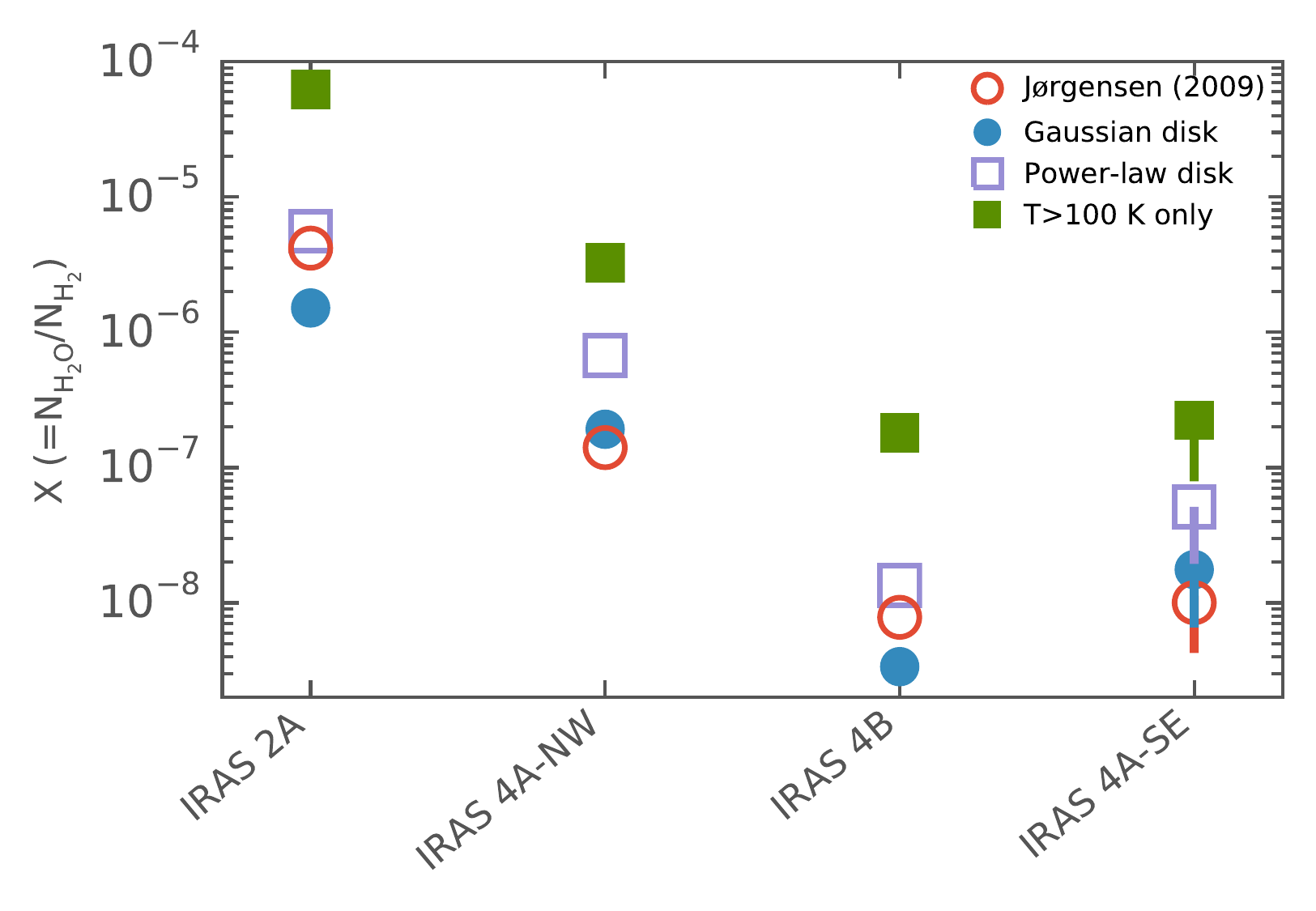}
        \caption{ Relative abundance of gas-phase water derived from observations 
        of p-H$_2^{18}$O by \citep{persson12} and continuum (for estimating 
        the H$_2$ abundance) for different methods of estimating the disk gas 
        mass. The values of the mass by \citet{jorgensen09} used by     
        \citet{persson12} are drawn as red open circles. The size of the symbols 
        reflects the uncertainty; for IRAS~4A-SE the values are upper-limits. }
        \label{fig:water_frac}
\end{figure}

\citet{persson12} estimate gas phase water abundances to total 
(cold+warm) H$_2$ densities for four of the sources observed in this 
study by using masses derived by \citet{jorgensen09} from the 850~$\mu$m 
continuum in the same beam. In \figref{fig:water_frac} these values, and
new estimates derived using the masses from this study are shown. The largest 
difference is seen for IRAS~4B and the upper limit of IRAS~4A-SE. 
With the new total disk/compact masses the warm-water abundance is 
slightly different. However, it is expected that  the whole 
structure will not have temperatures above 100~K, thus the fractional abundance 
using the total mass inferred from the power-law and Gaussian disk can be seen 
as a lower limit to the relative gas phase abundance of water. 

Correcting for the  amount of material in the disk above 100~K increases the 
fractional abundances  with up to almost two orders of magnitude. However, as 
shown by \citet{harsono15b}, part of the warm water emission could originate in 
the envelope surrounding the disk,  i.e., the hot corino, thus making the 
fractional abundances for $T>100$~K material upper limits. This  part of the 
model is not captured in the modeling of the disk,  which focuses on the cold 
dense region. We note that the power-law disk model does not include any vertical
structure. A temperature gradient along the vertical direction of the
disk could increase the amount of warm gas ($T>100$~K), thus decreasing the 
abundance of gas phase water.

Given the models used in this study, correcting for the material with $T>100$~K 
gives relative water abundances as high as $6.2\times10^{-5}$ for IRAS~2A, 
comparable to what is expected from sublimated ices ($10^{-4}$). For the rest of 
the sources studied, the same abundances are lower by $1\sim3$ orders of 
magnitude from this high value in IRAS~2A. The disk in IRAS~4B is compact and 
dense (flat profile and high surface density), this causes the very low 
relative abundance of water. The envelope power-law is also flatter than the 
other sources. These results show that disks are not as dry as 
previously stated, but are not ``wet'' either.

\begin{table}[ht]
\caption{Derived relative water abundance ($\times10^{-7}$) for the various 
models shown in \figref{fig:water_frac}. A corresponds to abundance derived from 
\citet{jorgensen09} (red open circle, \figref{fig:water_frac}), B from the 
Gaussian disk derived in this study (blue filled circle), C from the power-law 
disk (open purple square), and D from material in the power-law disk with only 
with T>100~K (green filled square).}\label{tab:water_abund}
\centering
\begin{tabular}{l c c c c}
\hline\hline          
Source     & A & B & C &  D \\
\hline
   IRAS 2A              & $42$   & $15$   &  $56$   & $620$ \\
   IRAS 4A-NW   & $1.4$  & $2$    &  $6.76$ &  $33$ \\
   IRAS 4B              & $0.1$  & $0.03$ &  $0.1$  &  $1.8$ \\
   IRAS 4A-SE   & $<0.1$ & $<0.2$ &  $<0.5$ & $<2$ \\
\hline
\end{tabular}
\tablefoot{The typical uncertainty is $20\sim30$\%.}
\end{table}

\section{Summary and outlook}
The deeply embedded nature of Class~0 sources adds to the difficulty of 
determining the physical structure on small, sub-300~AU, scales. Several 
parameters need to be determined. On small scales, a thin-disk model can 
approximate the emission and make it possible to constrain the radial density 
profile, assuming a fixed temperature distribution. For the five Class~0 sources 
studied here, the derived disk radii are similar ($90-170$~AU), and masses from 
fitting a power-law disk range from $9-140\times10^{-3}$~\Msun. Most of the 
derived power-law disk masses agree with previous estimates using other methods.
The gas surface density varies between $2.4-51$~g~cm$^{-2}$ (at 
$r_0=50$~AU). The temperature and density profiles have a smooth transition to 
the envelope in general. 

The inferred disk/compact masses are high, comparable to or higher than the assumed 
stellar mass of 0.05~\Msun. VLA~1623 is the only source in this study 
where a stellar mass of 0.2~\Msun\ has been determined. The ratio between the 
disk and 
stellar mass determines when disk instabilities may develop. At ratios higher 
than 0.25 the disk becomes prone to gravitational instabilities 
\citep[e.g.,][]{lodato04,dong15}, which would apply to most of our disks 
up to $M_*\approx0.3$\Msun.

The fractional amount of material above 100~K in the whole disk varies 
between $9-30$~\%, with an assumed temperature profile of $q=0.5$ (see 
Sect.~\ref{sec:plawmodel} for choice of $q$). A lower $q$ would increase the 
percentages, and a higher $q$ decrease. The main assumption is that the compact 
structure is indeed a disk that can be 
described with this power-law model. Using 
previous observations of p-H$_2^{18}$O, we estimate 
relative gas phase water abundances to total H$_2$ densities and for 
H$_2$ warmer than 100~K. The relative water abundance in the warm gas 
($>$100~K) of these disks are $6.2\times10^{-5}$ (IRAS~2A), 
$0.33\times10^{-5}$ (IRAS~4A-NW), $1.8\times10^{-8}$ (IRAS~4B), and 
$<2\times10^{-8}$ (IRAS~4A-SE). Thus, the gas-phase water abundance can be as 
high as the expected value for sublimated ice of $10^{-4}$ (as in IRAS~2A), but 
are lower for the other sources studied.

The techniques developed here can be applied to large samples of sources.
The next step is to image both the continuum and molecular line
tracers at higher angular resolution with ALMA and with good $uv$
coverage on all scales to constrain  any disk-like structure as
well as the transition to the larger scale envelope. 

\begin{acknowledgements}
  We wish to thank the IRAM staff, in particular Arancha
  Castro-Carrizo and Chin Shin Chang, for their help with the
  observations and reduction of the data. IRAM is supported by
  INSU/CNBRS(France), MPG (Germany), and IGN (Spain). Fruitful
  discussions with Steven Doty are acknowledged. This research made
  use of Astropy, a community-developed core Python package for
  Astronomy \citep{astropy13}. MVP and EvD acknowledge EU A-ERC grant
  291141 CHEMPLAN and a KNAW professorship prize. JJT acknowledges
  support from grant 639.041.439 from the Netherlands Organisation for
  Scientific Research (NWO). DH is funded by Deutsche Forschungsgemeinschaft Schwerpunktprogramm (DFG SPP 1385) The First 10 Million Years of the Solar System – a Planetary Materials Approach. 
  JKJ acknowledges support from a Lundbeck Foundation 
  Group Leader Fellowship as  well as the European Research Council (ERC) under 
  the European Union’s Horizon 2020 research and innovation programme (grant 
  agreement No 646908) through ERC Consolidator Grant “S4F”. Research at Centre 
  for Star and Planet Formation is funded by the Danish National Research 
  Foundation. This work has benefited from research funding from the European 
  Community's sixth Framework Programme under RadioNet R113CT 2003 5058187. This 
  paper made use of the following ALMA data: ADS/JAO.ALMA 2013.1.01004.S. ALMA 
  is a partnership of ESO (representing its member states), NSF (USA), and NINS 
  (Japan), together with NRC (Canada) and NSC and ASIAA (Taiwan), in cooperation 
  with the Republic of Chile. The Joint ALMA Observatory is operated by ESO,
  AUI/NRAO, and NAOJ.  The Submillimeter Array is a joint project
  between the Smithsonian Astrophysical Observatory and the Academia
  Sinica Institute of Astronomy and Astrophysics and is funded by the
  Smithsonian Institution and the Academia Sinica. S.P.L. acknowledges
  support from the Ministry of Science and Technology of Taiwan with
  Grants MOST 102-2119-M-007-004- MY3.  The authors wish to recognize
  and acknowledge the very significant cultural role and reverence
  that the summit of Mauna Kea has always had within the indigenous
  Hawaiian community. We are most fortunate to have the opportunity to
  conduct observations from this mountain.
\end{acknowledgements}

\bibliographystyle{aa} 
\bibliography{bibfile}

\clearpage
\begin{appendix}

\section{Output from the various analysis steps}\label{app:steps}
\begin{figure*}[ht]
        \centering
        \includegraphics[width=0.75\linewidth]{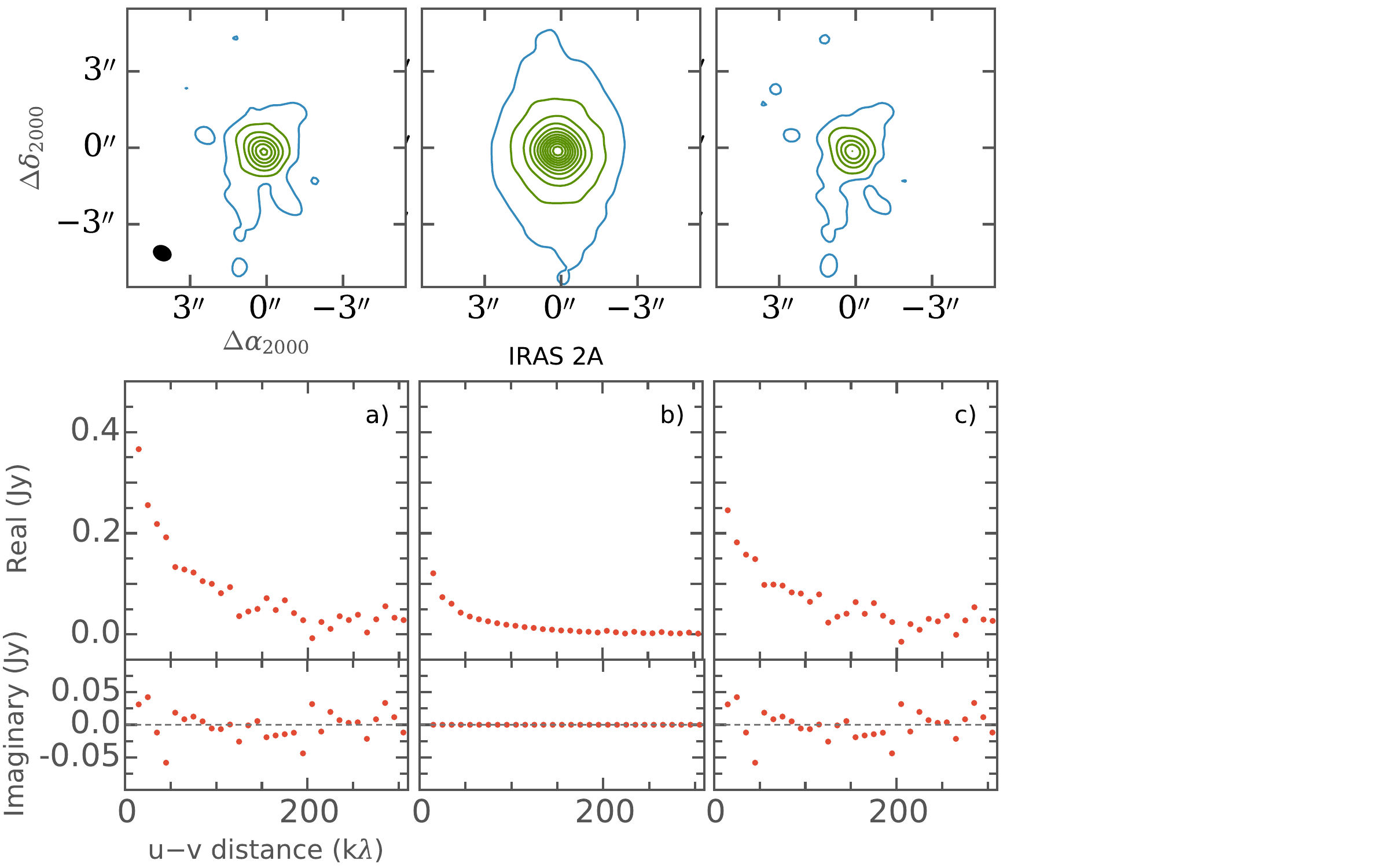}
        \caption{Images and visibilities of data preparation steps for IRAS~2A. From left to right: raw data, envelope model, raw data with envelope subtracted. The images start at 3$\sigma$, in steps of 3$\sigma$ until 9$\sigma$ where it is in steps of 9$\sigma$, $\sigma$=1.6~mJy ($\sigma$=0.2~mJy for envelope model image).}
        \label{fig:steps_i2a}
\end{figure*}

\begin{figure*}[ht]
        \centering
        \includegraphics[width=0.75\linewidth]{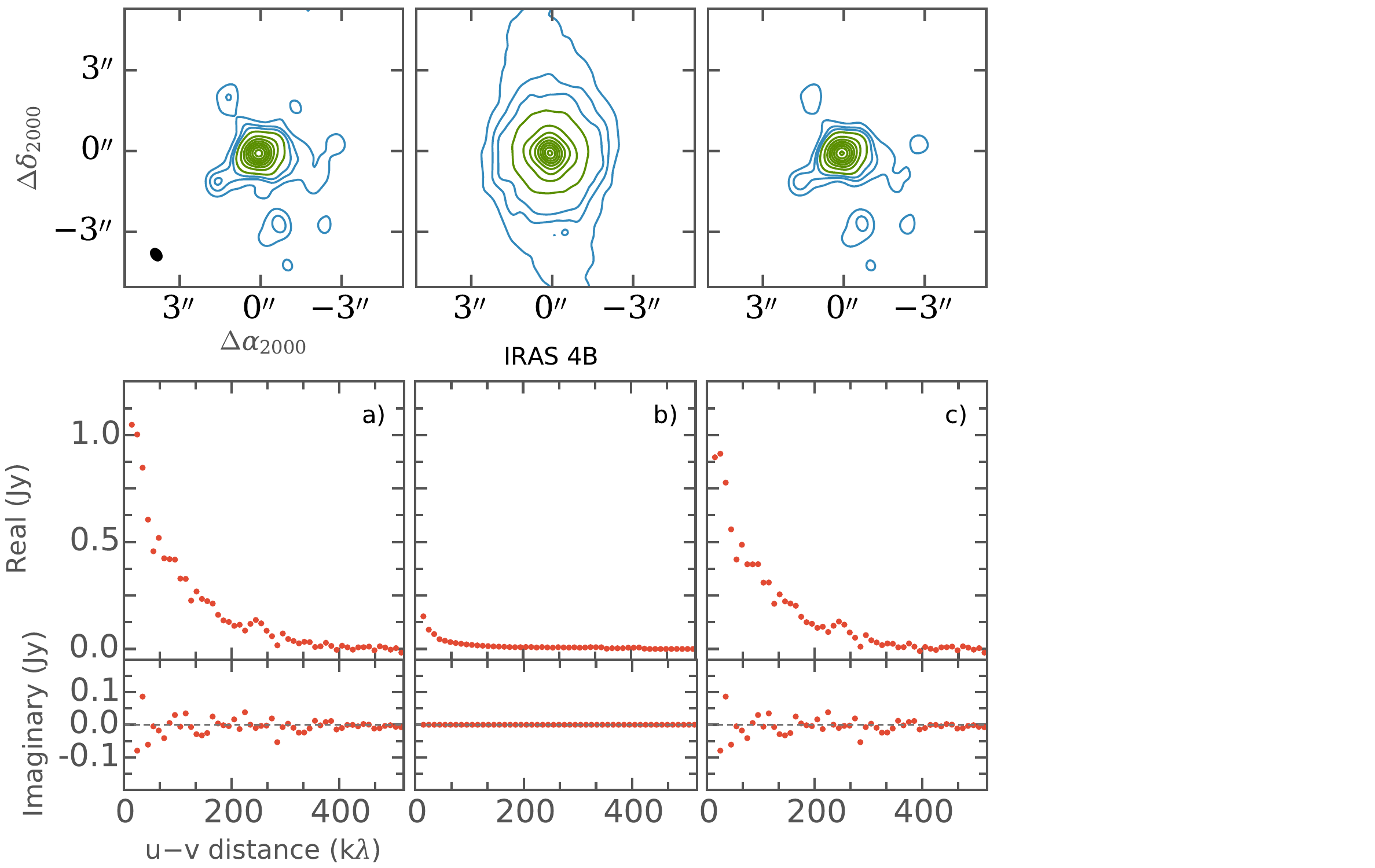}
        \caption{Images and visibilities of data preparation steps for IRAS~4B. From 
        left to right: raw data, envelope model, raw data with envelope subtracted. 
        The images start at 3$\sigma$, in steps of 3$\sigma$ until 15$\sigma$ where 
        it is in steps of 15$\sigma$, $\sigma$=2~mJy ($\sigma$=0.13~mJy for envelope 
        model image). As pointed out in the text, the observations of IRAS 
        4B are dynamic range limited, which causes the compact emission around the 
        main continuum peak seen in the images above; these are convolution 
        artifacts and are not real. }
        \label{fig:steps_i4b}
\end{figure*}

\begin{figure*}[ht]
        \centering
        \includegraphics[width=0.75\linewidth]{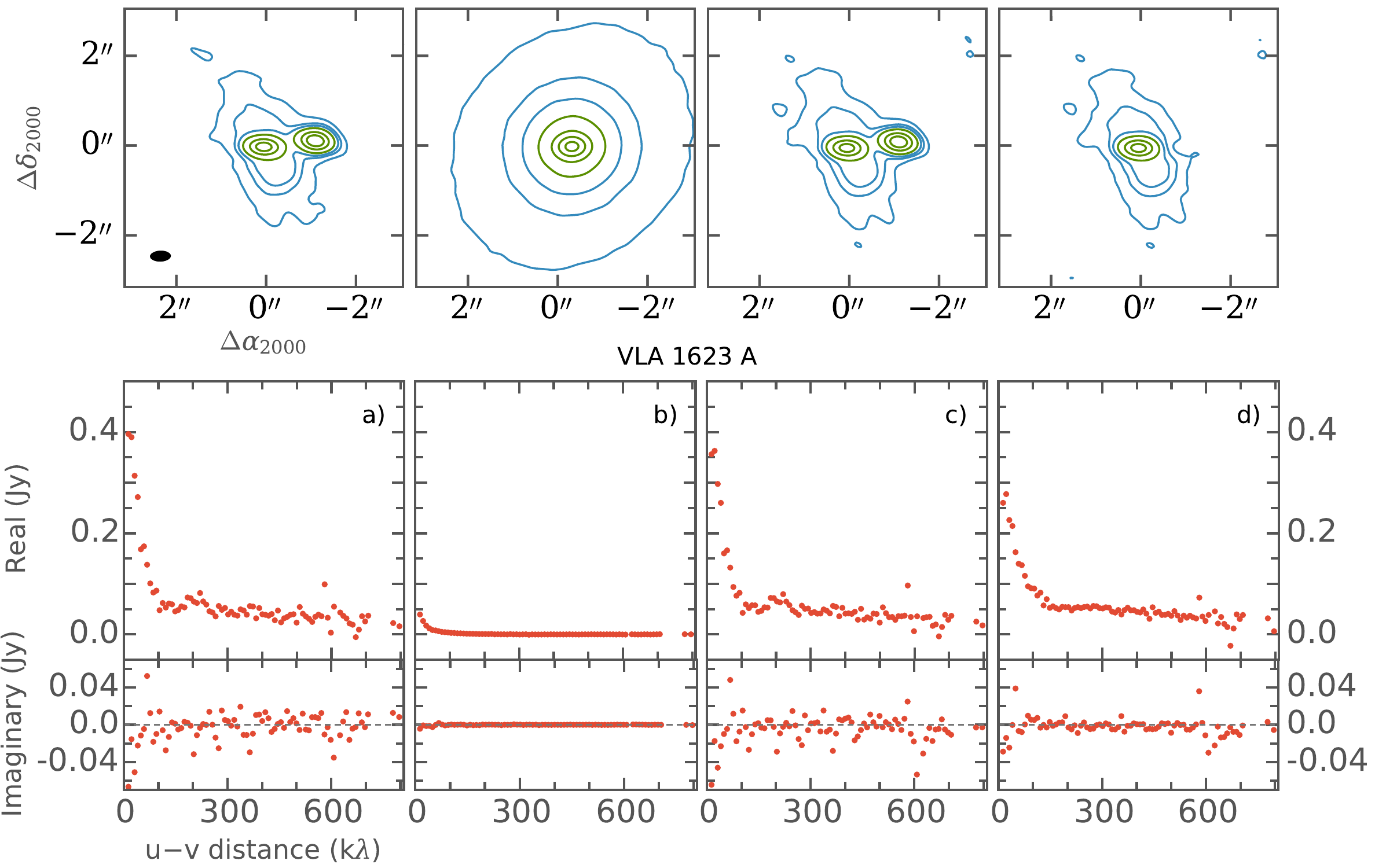}
        \caption{Images and visibilities of data preparation steps for VLA~1623~A. Top: imaged data; bottom: binned real and imaginary visibilities. From left to right; a) raw data, b) envelope model, c) raw data with envelope subtracted, d) raw data with envelope and companion subtracted. The contours start at 3$\sigma$ then insteps of 7$\sigma$ until 30$\sigma$ and then in steps of 30$\sigma$, $\sigma$=0.6~mJy (0.02~mJy for envelope model 
        image). }
        \label{fig:steps_vla}
\end{figure*}

The output (image, real, and imaginary visibility amplitude) of each step 
of the data preparation,  i.e., envelope and companion subtraction, are shown in 
Figures~\ref{fig:steps_i2a} to \ref{fig:steps_vla}. In each figure, the raw 
data, envelope model, envelope subtracted data, and companion subtracted data are 
shown. For VLA~1623~A the companion subtraction makes a clear difference, 
which can be seen even in the visibilities.


\section{Working with visibilities}

\subsection{Binning visibilities - vector averaging}

The visibilities are binned in annuli around the origin of the \uv\ plane.
In the following equations, $N_p$ is the number of points in the bin, $re_i$ the 
real part of each visibility, and $Re$ the binned real amplitude in each bin; the 
same notation holds for $im_i$ and $Im$:
\begin{eqnarray}
        Re = \dfrac{\sum^{N_p}_{i=0}re_i}{N_p} ; Im = 
        \dfrac{\sum^{N_p}_{i=0}im_i}{N_p}\\
        \sigma_{Re} = \sqrt{\dfrac{\sum_{i=0}^{N_p} re^2_i - N_p Re^2}{N_p -1 }} ;  
        \sigma_{Im} = \sqrt{\dfrac{\sum_{i=0}^{N_p} im^2_i - N_p Im^2}{N_p -1 }}
\end{eqnarray}
The combined amplitude is simply the square root of the sum of the squared real 
and imaginary amplitudes (i.e.,\ $A = \sqrt{Re^2+Im^2}$), the standard deviation 
is then the error propagation of the individual errors of the real and imaginary 
parts.
\begin{equation}
        \sigma_A = \sqrt{ \dfrac{\left(\dfrac{Re \sigma_{Re}}{A}\right)^2 + 
        \left(\dfrac{Im \sigma_{Im}}{A}\right)^2 }{N_p - 2}}
\end{equation}

\subsection{Rotation and inclination}\label{app:orientation}

Each \emph{u} and \emph{v} coordinate is rotated $PA$ degrees and inclined $i$ 
degrees. This is accomplished by first calculating the \uv\ distances from the 
origin
\begin{equation}
         r_{uv}=\sqrt{u^2+v^2} 
\end{equation}
 and the angle of the new point by subtracting the position angle from the 
 current direction of the point (measured east of north):
\begin{equation}
         \gamma = \arctan\left(\frac{v}{u}\right) - PA
.\end{equation}
Calculating the rotated and inclined (along the rotated \emph{u}-axis) 
coordinate system, we get new \emph{u} and \emph{v} coordinates
\begin{eqnarray}
         u^\prime = r_{uv} \sin \gamma \cos i\\
         v^\prime = r_{uv} \sin \gamma 
\end{eqnarray}
with the new \uv\ distance naturally given by $r^\prime_{uv} = \sqrt{u^{\prime 
2} + v^{\prime 2}}$.

\section{Envelope-only results}
This section presents the results of the envelope-only fitting of the 
interferometric visibilities. As shown in Section~\ref{sec:discussion}, 
these models do not necessarily fit the large-scale emission. For 
IRAS~4A-SE two fits were performed, one including all visibilities and one 
with only visibilties with \uv\ distances shorter than 150~\klam\ to minimize 
interference from the binary source.
\begin{figure*}[ht] 
        \label{fig:env_fit_all}
        \begin{minipage}[b]{0.45\linewidth}
                \centering
                \includegraphics[width=.9\linewidth]{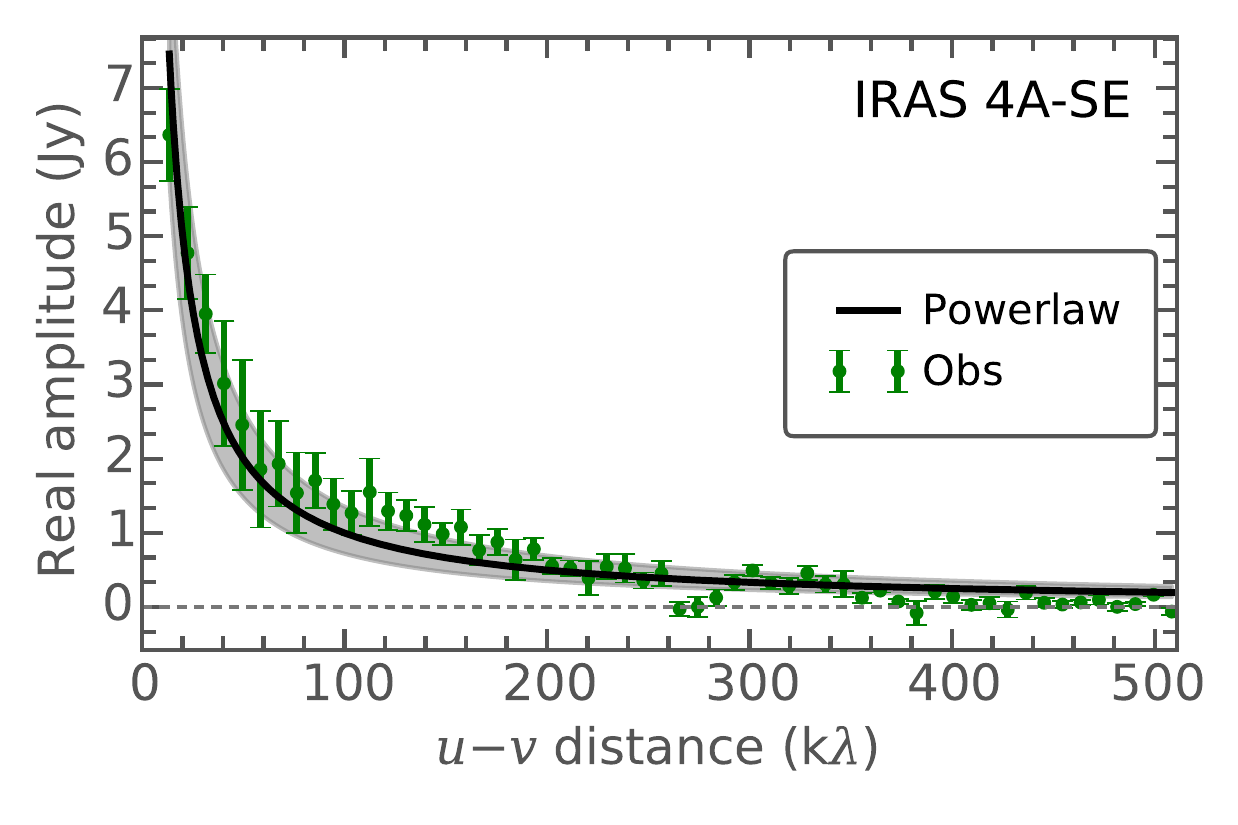} 
                \caption{Power-law envelope model fit to IRAS~4A-SE data. All the \uv\ 
                distances were fit.} 
                \vspace{2ex}
        \end{minipage}
        \hfill
        \begin{minipage}[b]{0.45\linewidth}
                \centering
                \includegraphics[width=.9\linewidth]{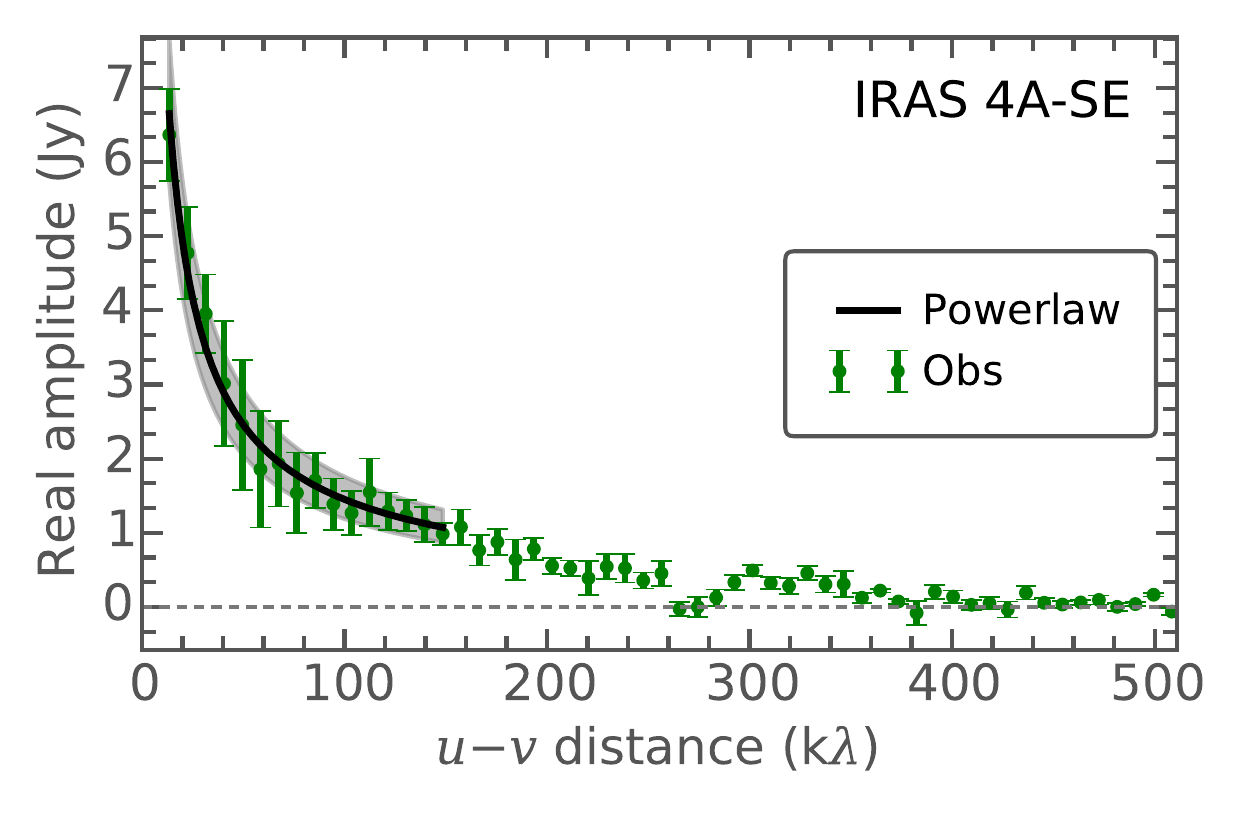} 
                \caption{Power-law envelope model fit to IRAS~4A-SE data. Only \uv\ 
                distances shorter than 150~\klam\ were fit to make sure the model is 
                not contaminated by the companion source. The parameters for the fit are 
                $C=48\pm4$ and $a = -0.76\pm0.02$. 
                }
                \vspace{2ex}
        \end{minipage}
        \begin{minipage}[b]{0.45\linewidth}
                \centering
                \includegraphics[width=0.9\linewidth]{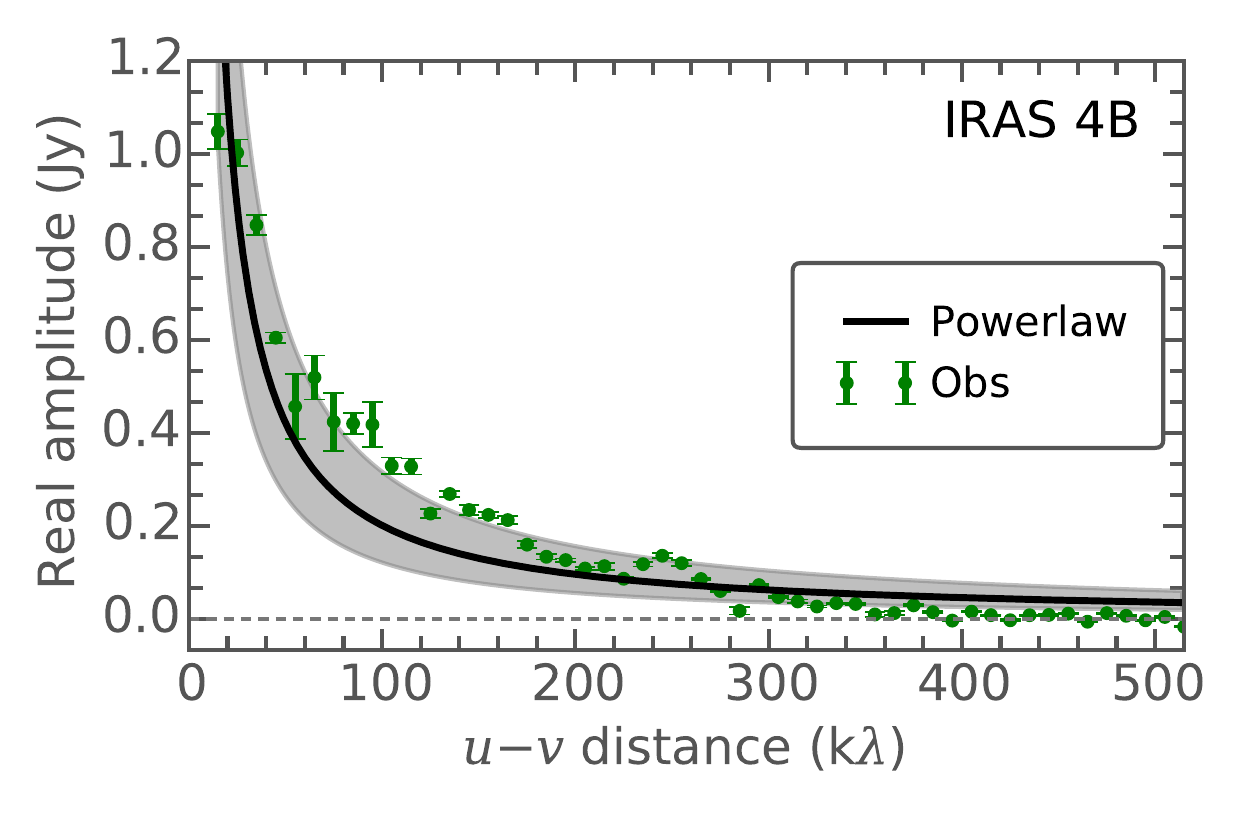} 
                \caption{Power-law envelope model fit to IRAS~4B.} 
                \vspace{2ex}
        \end{minipage}
        \hspace{8ex}
        \begin{minipage}[b]{0.45\linewidth}
                \centering\
                \hfill
                \includegraphics[width=.9\linewidth]{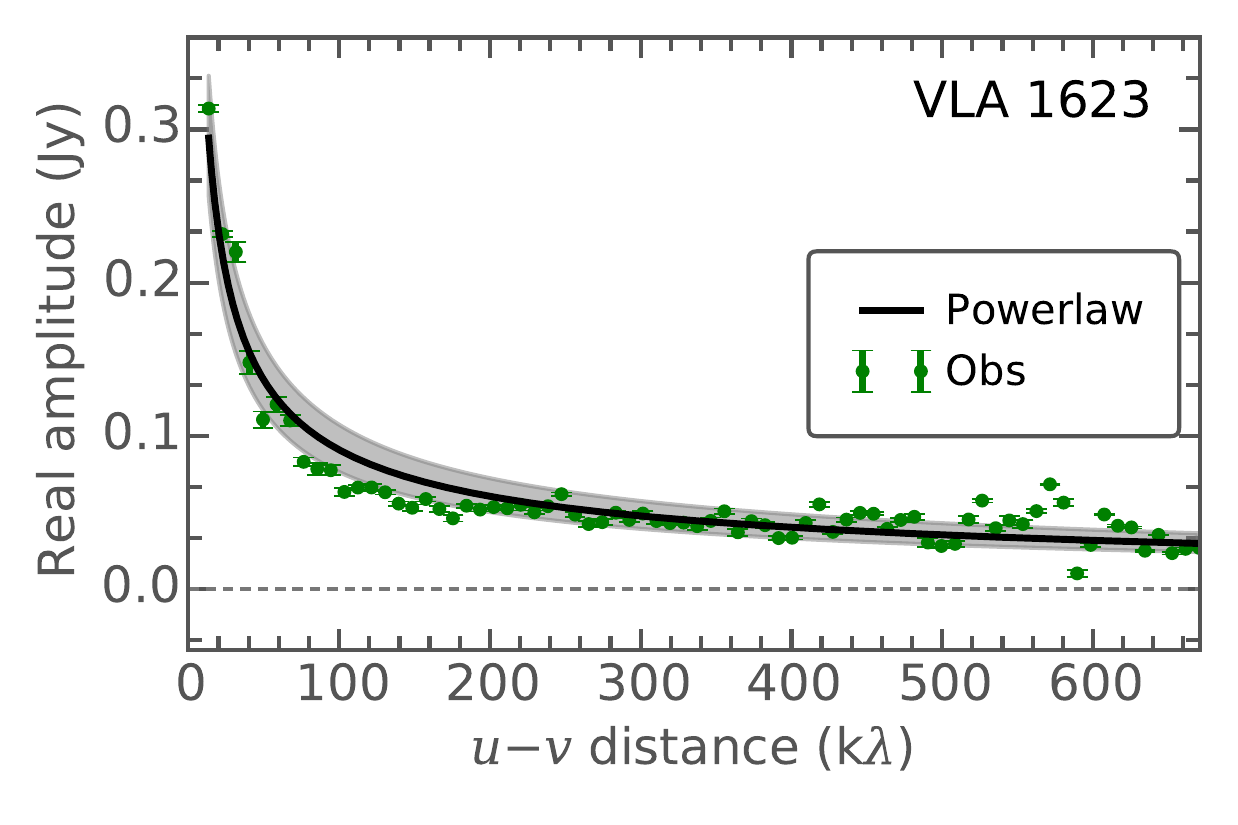} 
                \caption{Power-law envelope model fit to VLA~1623.} 
                \vspace{2ex}
        \end{minipage} 
        
\end{figure*}

\begin{table}[ht]
\caption{Parameters for the envelope-only models from a fit 
$V(b)=Cb^{a}$.}\label{tab:env_only}
\centering
\begin{tabular}{l c c }
\hline\hline          
Source                  & $C$                   & $a$ \\
\hline                       
   IRAS 2A              & $3.2\pm0.4$   & $-0.78\pm0.03$ \\
   IRAS 4B              & $28\pm6$              & $-1.07\pm0.05$ \\
   IRAS 4A-SE   & $102\pm15$    & $-1.01\pm0.04$ \\
   VLA 1623     & $1.4\pm0.1$   & $-0.59\pm0.02$ \\
\hline
\end{tabular}
\end{table}

\section{Envelope parameters}
\begin{table}[ht]
\caption{Parameters used for the envelope models.}\label{tab:envmodels}
\centering
\begin{tabular}{l c c c c}
\hline\hline                 
Source & $p_\mathrm{env}$ & $r_\mathrm{in}$ [AU] & $r_\mathrm{out}$ [AU] & 
$n_\mathrm{in}$ (cm$^{-3}$)  \\
\hline                        
   IRAS 2A & 1.7 & 35.9 & 17950 & $4.9\times 10^8$ \\
   IRAS 4A & 1.8 & 33.5 & 33500 & $3.1\times 10^9$ \\
   IRAS 4B & 1.4 & 15.0 & 12000 & $2.0\times 10^9$ \\
   VLA 1623& 1.4 & 4.3  & 10320 & $1.6\times 10^9$ \\
\hline
\end{tabular}
\tablefoot{
From \citet{kristensen12,jorgensen02,murillo13a,murillo13b} and references 
therein.
}
\end{table}

\end{appendix}

\end{document}